\documentclass[structabstract]{aa}  
\usepackage{natbib} 
\bibpunct{(}{)}{;}{a}{}{,} 

\usepackage{graphicx}
\usepackage{txfonts}
\def\farcs{\hbox{$.\!\!^{\prime\prime}$}}
\def\degr{\hbox{$^\circ$}}
\def\arcsec{\hbox{$^{\prime\prime}$}}

\newcommand{\sii}{[\ion{S}{ii}]}
\newcommand{\azii}{[\ion{N}{ii}]}
\newcommand{\oi}{[\ion{O}{i}]}
\newcommand{\cii}{[\ion{C}{ii}]}

\newcommand{\ho}{H$_2$O}

\newcommand{\kms}{km\,s$^{-1}$}
\newcommand{\um}{$\mu$m}
\newcommand{\lam}{$\lambda$}

\newcommand{\wm}{W\,m$^{-2}$}
\newcommand{\cmc}{cm$^{-3}$}
\newcommand{\ang}{$\AA$}

\newcommand{\macc}{$\dot{M}_{\rm acc}$}
\newcommand{\mjet}{$\dot{M}_{\rm jet}$}

\newcommand{\msolyr}{M$_{\odot}$\,yr$^{-1}$} 
\newcommand{\msol}{M$_{\odot}$} 
\newcommand{\lsol}{L$_{\odot}$} 
\newcommand{\be}{\begin{equation}} 
\newcommand{\ee}{\end{equation}} 
\begin{document}
   \title{{\it Herschel}/PACS observations of young sources in Taurus: \\
     the far-infrared counterpart of optical jets\thanks{{\it Herschel} is an ESA space observatory with science instruments provided by European-led Principal Investigator consortia and with important participation from NASA.}}


   \author{L. Podio
          \inst{1,2}
          \and
	  I. Kamp
          \inst{2}
          \and
          D. Flower
          \inst{3}
          \and
          C. Howard
          \inst{4}
          \and
          G. Sandell
          \inst{4}
          \and
          A. Mora
          \inst{5}
          \and
          G. Aresu
          \inst{2}
          \and
          S. Brittain
          \inst{6}
          \and
          W. R. F. Dent
          \inst{7}
          \and
          C. Pinte
          \inst{1}
          \and
          G. J. White
          \inst{8,9}
          }

   \institute{
     Institut de Plan\'etologie et d'Astrophysique de Grenoble, 414, Rue de la Piscine, 38400 St-Martin d'H\`eres, France  
     \email{linda.podio@obs.ujf-grenoble.fr}
   \and 
   Kapteyn Institute, Landleven 12, 9747 AD Groningen, The Netherlands 
   \and 
   Physics Department, The University of Durham, Durham DH1 3LE, United Kingdom 
   \and
   SOFIA-USRA, NASA Ames Research Center, USA  
   \and
   ESA-ESAC Gaia SOC, PO Box 78, 28691 Villanueva de la Ca\~{n}ada, Madrid, Spain  
   \and
   Department of Physics \& Astronomy, 118 Kinard Laboratory, Clemson University, Clemson, SC 29634, USA  
   \and
   ALMA, Avda Apoquindo 3846, Piso 19, Edificio Alsacia, Las Condes, Santiago, Chile  
   \and
   Department of Physical Sciences, The Open University, Milton Keynes MK7~6AA, United Kingdom  
   \and
   RALSpace, The Rutherford Appleton Laboratory, Chilton, Didcot, OX11 0QX, UK  
             }

   \date{Received; accepted}

 
  \abstract
   {Observations of the atomic and molecular line emission associated with jets and outflows emitted by young stellar objects provide sensitive diagnostics of the excitation conditions, and can be used to trace the various evolutionary stages they pass through as they evolve to become main sequence stars.}
   {To understand the relevance of atomic and molecular cooling in shocks, and how accretion and ejection efficiency evolves with the evolutionary state of the sources, we will study the far-infrared counterparts of bright optical jets associated with Class I and II sources in Taurus (T Tau, DG Tau A, DG Tau B, FS Tau A+B, and RW Aur).}
   {We have analysed {\it Herschel}/PACS observations of a number of atomic (\oi63\um, 145\um, \cii158\um) and molecular (high-J CO, \ho, OH) lines, collected within the Open Time Key project GASPS (PI: W.~R.~F. Dent). To constrain the origin of the detected lines we have compared the obtained FIR emission maps with the emission from optical-jets and millimetre-outflows, and the measured line fluxes and ratios with predictions from shock and disk models.}
   {All of the targets are associated with extended emission in the atomic lines; in particular, the strong \oi~63~\um\, emission is correlated with the direction of the optical jet/mm-outflow. The line ratios suggest that the atomic lines can be excited in fast dissociative J-shocks occurring along the jet.
       The molecular emission, on the contrary, originates from a compact region, that is spatially and spectrally unresolved, and lines from highly excited levels are detected (e.g., the o-\ho\,8$_{18}$ - 7$_{07}$ line, and the CO J=36-35 line). Disk models are unable to explain the brightness of the observed lines (CO and \ho\, line fluxes up to 10$^{-15}$-6~10$^{-16}$ \wm).  Slow C- or J- shocks with high pre-shock densities reproduce the observed \ho\, and high-J CO lines; however, the disk and/or UV-heated outflow cavities may contribute to the observed emission.}
   {Similarly to Class 0 sources, the FIR emission associated with Class I and II jet-sources is likely to be shock-excited. While the cooling is dominated by CO and \ho\, lines in Class 0 sources, \oi\, becomes an important coolant as the source evolves and the environment is cleared. The cooling and mass loss rates estimated for Class II and I sources are one to four orders of magnitude lower than for Class 0 sources. This provides strong evidence to indicate that the outflow activity decreases as the source evolves. 
}


   \maketitle
%
\section{Introduction}

Theoretical models \citep[e.g., ][]{shu94,konigl00} predict a tight correlation between the accretion of matter onto a young star and the ejection in winds and/or jets. 
Measurements of stellar accretion and mass loss \citep[e.g., ][]{hartigan95} support the general picture presented by these models, but the uncertainties of these measurements are too large to provide a quantitative test of the predictions of the ratio of the mass accretion rate to the mass loss rate.
Sources in the earliest stages in their evolution (Class 0) are not visible, and are often  indirectly identified by means of their strong ejection activity, which is manifested in the form of bipolar parsec-scale molecular
outflows often observed at millimetre wavelengths \citep[e.g., ][]{bachiller96}.
The ejection associated with evolved, optically visible T Tauri stars (i.e., Class II) is instead usually traced by bright blue- and red- shifted forbidden emission lines present at optical and near-infrared (NIR) wavelengths \citep[e.g., ][]{hartigan95}.
For Class I sources and some Class II sources both the molecular outflow and the optical jet have been observed \citep{gueth99,pety06}. 
These observations show that the two components are connected. 
The optical/NIR forbidden lines trace hot ($\sim$10$^4$ K) atomic gas, which is believed to have been extracted from the disk, and accelerated in
the observed fast and collimated {\it jets} (velocities up to hundreds of
\kms\, and jet widths smaller than 200 AU). 
The millimetre observations, instead, trace cold ($\sim$10-100 K) and slow (tens of \kms) gas, which is thought to be ambient gas that has been set into motion by the jet propagation (i.e., {\it jet-driven molecular outflow},  e.g., \citealt{raga93,cabrit97}). However, collimated high velocity molecular gas (velocity up to $\sim$60 \kms, \citealt{lefloch07}) has also previously been detected at millimetre wavelengths, questioning this simple picture, and suggesting that molecules can also be extracted from the disk and accelerated in the jet \citep{pontoppidan11,panoglou12}.\\

In this context, observations at far-infrared wavelengths allow us to trace the intermediate {\it warm gas component} in the jet/outflow system.
Previous observations from the Infrared Space Observatory (ISO, \citealt{kessler96}) targeting outflow sources have shown emission in a large number of atomic (\oi, \cii) and molecular (\ho, CO, OH) lines.
Despite the very low spectral and spatial resolution offered by ISO, analysis of the line fluxes and ratios indicates that the bulk of the detected \oi\, and molecular emission is most likely to be excited in the shocks occurring along the jet/outflow \citep{nisini96,nisini99,giannini01}. The line fluxes have previously been used to estimate the cooling in the atomic and molecular lines, and to quantify the outflow efficiency as the ratio between the total luminosity radiated away in the far-infrared lines, L~(FIR), and the source bolometric luminosity, L$_{bol}$ \citep{giannini01, nisini02}. 
However, because of its limited sensitivity, ISO observations have been restricted to studies of bright and extended outflows from Class 0 and I objects. 
The ESA {\it Herschel} Space Observatory ({\it Herschel}) has allowed, for the first time, observations of the FIR counterparts of optical jets associated with Class I and Class II sources whose environment has been largely cleared.

As the source evolves, the accretion/ejection activity is expected to decrease, with the surrounding cloud material being either accreted onto the star, or dispersed by the jet. As a consequence, the optical jet will become visible while the emission at far-infrared wavelengths should be expected to be fainter and less extended than in Class 0 sources. 
However, FIR observations of the ejection activity associated with more evolved Class I and II sources is interesting for the following reasons:
\begin{itemize}
\item[-] The source, disk, and accretion properties of T Tauri stars are well-known, and so we can study the correlation between the detected ejection phenomena and these properties. Thus, we can estimate the mass ejection to mass accretion ratio;
\item[-] It is usually the case that Class 0 sources are observable only at millimetre wavelengths, whilst the ejection activity from T Tauri stars is detected only in the optical. FIR lines can however be detected from Class 0 to Class II sources, therefore facilitating a way to form an evolutionary picture of jet activity;
\item[-] The detection of molecular emission in sources whose environment has been cleared may support the idea of a disk-wind molecular component providing strong constraints to existing models of jet launching \citep{panoglou12}.
\end{itemize}

Thus we have analysed the FIR emission from five Class I and II sources in Taurus (d$\sim$140 pc).
These sources were observed as part of the {\it Herschel} Open Time Key project GASPS ({\it GAS in Protoplanetary Systems}, PI: W. R. F. Dent) using the PACS integral field spectrometer \citep{poglitsch10} on board {\it Herschel} \citep{pilbratt10}.
These PACS observations provide maps of the emission in a number of atomic and molecular lines  (see, \citealt{mathews10} and Dent et al., {\it in preparation}). 
The five sources presented in this paper form a subset of the Taurus sample analysed in Howard et al., {\it in preparation}.
These sources were selected on the basis of their association
with bright and extended stellar jets detected in the typical \sii, \azii, and \oi\, optical forbidden lines \citep[e.g., ][]{hartigan95,hirth97}. 
{The details of the observations, and the applied data reduction processing, are described in Sect.~\ref{sect:obs}.
In Sect.~\ref{sect:results} we show the obtained spectra and maps and compare the spatial distribution of the atomic and molecular emission with that of the associated optical jets. 
In Sect.~\ref{sect:discussion} we will compare the observed line fluxes and ratios with predictions from both disk and shock models. 
Hence, we will use the observed line fluxes and the results from shock modelling to estimate the far-infrared cooling and the mass loss rate. The comparison with the values estimated for Class 0 and I sources will allow us to place Class II sources into an evolutionary picture.
Finally, in Sect.~\ref{sect:conclusions} we summarise our conclusions.

\section{Observations}
\label{sect:obs}

The observations analysed in this paper have been acquired using the {\it Herschel}-PACS integral field spectrometer. 
The criteria used to select the sources are explained in Sect.~\ref{sect:sample_selection}.
Information on the instrumental settings and on data reduction processing are given in Sect.~\ref{sect:data_reduction}, and then in Sect.~\ref{sect:ext_emission} we will explain the procedure to distinguish extended and unresolved emission as observed with the PACS spectrometer.

\subsection{Sample selection}
\label{sect:sample_selection}

\begin{table*}
\caption[]{\label{tab:source_list} Source properties: stellar and bolometric luminosity, SED class, and position angle of the associated optical jet.}
\centering
    \begin{tabular}[h]{c|cccccc|c|c}
      \hline\hline
Source                 & L$_{*}$          & L$_{bol}$                & Class  & PA$_{jet}$ & Ref \\
                           & (\lsol)             & (\lsol)                	  &             & (\degr)       & \\
\hline
T Tau (N, Sa+Sb)   & 7.3, -           & 15.5, 10               & II, I      & 180, 270 & (1)\\
DG Tau  A             & 3.2               & 6.36                     & II        & 226         & (2)\\
DG Tau B               & -                  & $\sim$1.1           &      I    & 115          & (3)\\ 
FS Tau (Aa+Ab, B) & 0.15+0.17, - & 1.4, $>$0.5        & II, I   & -, 55 	     & (4)\\
RW Aur (A+B)        & 1.7+0.4        &   3.2                         & II      & 120          & (5)\\
\hline 
      \end{tabular}
\tablebib{
(1) L$_{*}$ (only for T Tau N) by \citet{white01}, L$_{bol}$ by \citet{kenyon95}, Class by \citet{furlan06,luhman10}, PA$_{jet}$ by \citet{solf99};
(2) L$_{*}$ and Class by \citet{luhman10,rebull10}, L$_{bol}$ by \citet{kenyon95}, PA$_{jet}$ by \citet{mundt83};
(3) L$_{bol}$ average value of the range estimated by \citet{kruger11}, Class by \citet{luhman10,rebull10}, PA$_{jet}$ by \citet{mundt83}; 
(4) L$_{*}$, L$_{bol}$, and Class for FS Tau Aa+Ab by \citet{hartigan03}, \citet{kenyon95}, and \citet{luhman10}, L$_{bol}$ and Class for FS Tau B by \citet{stark06}, and \citet{luhman10,rebull10}, PA$_{jet}$ by \citet{mundt84}; 
(5) L$_{*}$ by \citet{white01}, L$_{bol}$ by \citet{kenyon95}, Class by \citet{furlan06,luhman10},  PA$_{jet}$ by \citet{hirth94,hirth97}. 
}
\end{table*}

Many of the sources in Taurus are associated with stellar jets based on the detection of extended emission in optical forbidden lines, such as \sii, \oi, and \azii\, lines in the 6300-6700 \ang\, wavelength range, which are thought to be excited in the shocks occurring along the jet \citep[see, e.g., ][]{hartigan95,hirth97}.
Within the Taurus sample (Howard et al., {\it in preparation}), we selected five well-known jet sources (T Tau, DG Tau A, DG Tau B, FS Tau A+B, and RW Aur) having the following characteristics:
\begin{itemize}
\item[-] they are associated with bright jet emission in optical forbidden lines, extending on spatial scales larger than the PACS spatial sampling (i.e., $\sim$9\farcs4);
\item[-] they show extended emission in the \oi~63~\um\, line (see Sect. \ref{sect:ext_emission} and \ref{sect:atomic} for details);
\item[-] they also show emission in the atomic \oi~145~\um\, and \cii~158~\um\, lines and in a number of \ho\, and high-J CO lines (see Sect. \ref{sect:results}) which make it possible to compare observed line ratios and fluxes with predictions from disks and shocks models (see Sect. \ref{sect:discussion}).
\end{itemize}
These characteristics make these sources ideal candidates to study the far-infrared atomic and molecular counterpart of optical stellar jets. 
Note that, with the exception of RW Aur,  
their environment is not completely cleared: T Tau, DG Tau B, and FS Tau B are associated with CO outflows detected at millimetre wavelengths \citep{edwards82,mitchell97,davis10}, and arc-shaped reflection nebulae associated with the optical jet have been observed for T Tau (W-E jet, blue lobe), DG Tau A (blue lobe), and DG Tau B (both lobes) by \citet{stapelfeldt97,stapelfeldt98}. This suggests that these nebulae are illuminated outflow cavities.

With the exception of T Tau, which has previously been observed in the FIR with ISO showing emission in \oi, \cii, OH, CO, and \ho\, lines \citep{spinoglio00}, and a tentative detection of the \oi~63~\um\, line at 3.5$\sigma$ for DG Tau A \citep{cohen88}, the sources in the sample have not previously been observed at FIR wavelengths. 

The stellar luminosity, bolometric luminosity, and Class of the selected sources are summarised in Table \ref{tab:source_list}, along with the position angle of the associated optical jets.
The class has been estimated  by means of recent Spitzer observations by \citet{furlan06,luhman10,rebull10} from the source spectral energy distribution (SED) in the infrared according to the classification of \citet{lada84,lada87}.
For multiple systems (T Tau, FS Tau, RW Aur) we list the  stellar, bolometric luminosity, and the class of each resolved component, along with the PA of the jet associated with each of them.
However, these systems, and their associated multiple jets are not resolved with PACS. Thus the continuum and line flux values quoted later in this paper 
refer to the whole system, as explained in the following section. 


\subsection{Instrumental setting and data reduction}
\label{sect:data_reduction}

\begin{table*}
  \caption[]{\label{tab:observations_log} Spectroscopic
  {\it Herschel}/PACS observations: order, arm, spectral coverage,
  resolution, integration times, and targeted lines.}
\centering
\begin{tabular}[h]{cccccc}
  \hline\hline
  Order & Arm & Spectral Coverage & R       & T$_{int}$ & Lines\\
            &        &             (\um)          & (\kms) &     (s)       & \\
   \hline
   3 & B & 62.93   - 63.43   & 88   & 1152           & \oi~$^3$P$_1$-$^3$P$_2$, o-\ho~8$_{18}$-7$_{07}$\\
   1 & R & 180.76 – 190.29 & 200  & 1152           & DCO$^{+}$ J=22-21\\
   \hline
   2 & B & 71.82   -  73.33  & 162 & 1592, 3184 & o-\ho~7$_{07}$-6$_{16}$, CH$^{+}$ J=5-4, CO J=36-35\\
   1 & R & 143.61 - 146.66 & 258 & 1592, 3184 & p-\ho~4$_{13}$-3$_{22}$, CO J=18-17, \oi~$^3$P$_0$-$^3$P$_1$ \\
   \hline
   2 & B & 78.37   - 79.73   & 147 & 1592, 3184 & o-\ho~4$_{23}$-3$_{12}$, p-\ho~6$_{15}$-5$_{24}$, OH~$^2 \Pi _{1/2, 1/2}$ - $^2 \Pi _{3/2, 3/2}$, CO J=33-32\\
   1 & R & 156.73 - 159.43 & 239 & 1592, 3184 & \cii~$^2$P$_{3/2}$-$^2$P$_{1/2}$, p-\ho~3$_{31}$-4$_{04}$\\
   \hline
   2 & B & 89.29   - 90.72   & 121 & 1592, 3184 & p-\ho~3$_{22}$-2$_{11}$, CH$^{+}$ J=4-3, CO J=29-28\\
   1 & R & 178.58 - 181.44 & 204 & 1592, 3184 & o-\ho~2$_{12}$-1$_{01}$, CH$^{+}$ J=2-1, o-\ho~2$_{21}$-2$_{12}$\\
   \hline 
 \end{tabular}
\end{table*}

The observations were acquired 
between February 2010 and March 2011.
The PACS integral field unit (IFU) 
allows simultaneous imaging of a
47\arcsec$\times$47\arcsec\, field of view, resolved into 5$\times$5
spatial pixels of  9.4\arcsec\,$\times$9.4\arcsec\, (also called {\em spaxels}). 
In spectroscopic mode for each spaxel a 1D spectrum is recovered
simultaneously in a selected spectral range in the blue
(B: 51-105 \um) and in the red (R: 102-220 \um) arm of the spectrometer.
The  observations were carried out in the chop-nod mode to remove
the background emission and with a single pointing on the source. 
Spectroscopic observations in line mode (PacsLineSpec) were obtained  by performing one nod cycle
for a total on-source integration time of 1152 s. These simultaneously covered
a selected wavelength range in the blue and in the red arms. 
Spectroscopic observations in range mode (PacsRangeSpec) were acquired by  performing 1 nod
cycle for T Tau, DG Tau A, and RW Aur, and 2 nod cycles for DG Tau B,
and FS Tau A+B. These  scanned three wavelength ranges in
the blue, and simultaneously, in the red arm, for a total integration time per
spectral range of 1592 s (1 nod cycle) and 3184 s (2 nod cycles).
In some of the observations: specifically in the case of DG Tau and DG Tau B,
the target was not centred on the central IFU spaxel, but at a position $\sim$6.7\arcsec\, away from
it. 
The order, arm, spectral coverage, spectral resolution, integration time, and targeted lines relative to the acquired  line and range spectra are summarised in Table \ref{tab:observations_log}.  
The observation identifiers (OBSIDs) are summarised in Table \ref{tab:obsid} in Appendix \ref{app:obs}.

All the data were reduced using HIPE 4.0.1467. 
The PACS pipeline included corrections for: saturated and bad pixel
removal, chop subtraction, relative spectral response function
correction, flat fielding, and mean of the two nods.
The spectra were Nyquist binned in wavelength, with non-overlapping bins set to
half the width of the instrumental resolution.
An aperture correction was not applied because (i)  in some cases the target was not
centred onto the central IFU spaxel and (ii) extended emission was
detected in some of the detected lines.
The continuum flux was recovered by summing the emission over 
the 5$\times$5 array, then by applying a first-order polynomial fit
to estimate the continuum level at the line rest wavelength.
The error on the continuum was calculated as the standard deviation of the difference between 
the estimated continuum level and the observed one in a region from 2 to 10
instrumental FWHM from the line rest wavelength. 
The line flux was recovered by summing the emission over those spaxels
where the line wass detected to avoid the suppression of faint lines by the noise in the outer spaxels.
The integrated flux of the detected lines was estimated by measuring the total flux under the gaussian line fit. We calculate the 1$\sigma$ flux by integrating a gaussian with height equal
to the continuum RMS, and width equal to the instrumental
FWHM; we report 3$\sigma$ upper limits for non-detections. 
All of the spectral ranges covered by our observations in the spaxel where the line emission is maximum are shown in Fig.~\ref{fig:lines_spectra} and 
the continuum and integrated line fluxes are summarised in Tables \ref{tab:cont_fluxes} and \ref{tab:line_fluxes} .

\subsection{Extended and unresolved emission}
\label{sect:ext_emission}

Fig. ~\ref{fig:OI_maps} (left panels) shows the 1D spectra in the 63.12 - 63.25~\um\, range for the 25 spaxels of the PACS array of the 5 sources in our sample. 
All of the maps show bright \oi~63~\um\, emission in a number of spaxels. This suggests that the line may originate from an extended region such as the jets detected at optical wavelengths, rather than from a circumstellar disk.
Indeed, if the spectroscopic point-spread-function (PSF) is smaller than the spaxel size, the emission from the source and the circumstellar disk (typical disk sizes in Taurus are $<$500 AU) will be spatially and spectrally unresolved with PACS, and thus confined in the central spaxel of the 5$\times$5 integral-field array.
On the contrary, if the emitting region is more extended than the spaxel size (9\farcs4$\times$9\farcs4) the line and continuum may be detected in the outer spaxels.

For the sources in our sample, this simple picture is complicated for two reasons:
(i) in  some observations the source is not centred in the central spaxel. In this case, even when observing an unresolved source, the emission is detected in a number of spaxels around the source position;
(ii) the spectroscopic PSF width in the observed spectral ranges is equal or larger than the spaxel size  (the PSF width for a source centred on the central spaxel vary between $\sim$9\arcsec\, at 60~\um\, and $\sim$13\arcsec\, at 180~\um), thus may cause line and continuum detection also in the outer spaxels. This effect is greater at the longer wavelengths, for bright sources, and when the source was not centred.

A simple method to distinguish between extended and non-extended line emission is to calculate the line-to-continuum ratio in each spaxel of the grid. This is defined as the ratio between the line flux, integrated over its spectral profile, and the continuum, integrated over one spectral resolution element.
If the line and the continuum emission originate from the same region, i.e. from the star-disk system, then the line and the continuum PSF will peak at the same position, and the line-to-continuum ratio is expected to be constant in all spaxels.
If the line originates from a region more extended and/or offset with respect to the continuum emitting region, we should measure higher line-to-continuum ratios in the outer spaxels along the direction of the extended emission.
Following this method, we compute the line-to-continuum ratios in all the spaxels for all the lines covered by our observations. 
In the \oi~63~\um\, maps shown in Fig.~\ref{fig:OI_maps} the presence of extended emission is indicated by line-to-continuum ratio which are larger than that measured on-source (i.e. where the continuum is at a maximum).

To quantify and localise the extended emission identified through the line-to-continuum ratios, we subtracted on-source line and continuum emission, which could be detected in the outer spaxels because of the spectroscopic PSF width  and/or off-centre sources.
This analysis was applied to all the obtained line emission maps and allowed us to determine which lines show evidence for extended emission, to identify the spaxels where this is detected and at which confidence level.
The mathematical formulation of the applied analysis is explained in  Appendix \ref{app:extended_emission}, while the obtained continuum and residual line emission contour plots for the \oi~63~\um\, line are shown in Fig.~\ref{fig:OI_maps} (right panels).

\begin{table*}
\begin{center}
\caption[]{\label{tab:cont_fluxes} Continuum emission, in Jy, in the spectral ranges covered by the {\it Herschel}/PACS observations. The continuum flux was estimated after integrating the emission over the 25 PACS spaxels. }
    \begin{tabular}[h]{c|ccccc}
\hline\hline
Source & T Tau & DG Tau A & DG Tau B & FS Tau A+B & RW Aur \\
\hline
Spectral range (\um) & \multicolumn{5}{c}{Cont $\pm \Delta$Cont (Jy)} \\
\hline
  62.93 -  63.43   &      132.4 $\pm$        0.8         &       14.6 $\pm$        0.3     &       11.9 $\pm$        0.2     &        5.2 $\pm$        0.2     &        0.7 $\pm$        0.2     \\
  71.82 -  73.33   &      129.9 $\pm$        0.1         &       15.5 $\pm$        0.1     &      11.48 $\pm$       0.06     &       4.65 $\pm$       0.05     &       2.25 $\pm$       0.10  \\   
  78.37 -  79.73   &      130.5 $\pm$        0.2         &       16.3 $\pm$        0.2     &       11.9 $\pm$        0.2     &        4.4 $\pm$        0.1     &        2.3 $\pm$        0.3     \\
  89.29 -  90.72   &      129.6 $\pm$        0.2         &       17.3 $\pm$        0.2     &       12.3 $\pm$        0.1     &       5.14 $\pm$       0.10     &        3.2 $\pm$        0.1     \\
 143.61 - 146.66   &     105.36 $\pm$       0.10         &      17.19 $\pm$       0.09     &      16.93 $\pm$       0.06     &       6.69 $\pm$       0.05     &       1.36 $\pm$       0.07\\     
 156.73 - 159.43   &      102.6 $\pm$        0.1         &       17.3 $\pm$        0.1     &      17.43 $\pm$       0.06     &       7.19 $\pm$       0.05     &        1.0 $\pm$        0.1     \\
 178.58 - 181.44   &       81.9 $\pm$        0.2         &       15.4 $\pm$        0.1     &      15.57 $\pm$       0.09     &       6.71 $\pm$       0.08     &        0.8 $\pm$        0.1     \\
\hline
     \end{tabular}
\end{center}
\end{table*}

\begin{table*}
\begin{center}
\caption[]{\label{tab:line_fluxes} Atomic and molecular line fluxes in \wm. The line fluxes were obtained by summing the emission over the spaxel where the line was detected (see text for details). }
    \begin{tabular}[h]{cc|ccccc}
\hline\hline
\multicolumn{2}{c}{Source} & T Tau & DG Tau A & DG Tau B & FS Tau A+B & RW Aur \\
\hline
Transition  & \lam (\um) & \multicolumn{5}{c}{F$_{line} \pm \Delta$F (\wm)} \\
\hline
     \oi\, $^3$P$_1$-$^3$P$_2$     &  63.184     &       1.91 $\pm$       0.01  10$^{-14}$   &      1.79 $\pm$       0.07  10$^{-15}$   &       7.3 $\pm$        0.4  10$^{-16}$   &        5.2 $\pm$        0.3  10$^{-16}$   &       2.0 $\pm$        0.2  10$^{-16}$   \\
      \oi\, $^3$P$_0$-$^3$P$_1$   & 145.525     &        8.8 $\pm$        0.3  10$^{-16}$   &       8.6 $\pm$        0.9  10$^{-17}$   &       2.4 $\pm$        0.2  10$^{-17}$   &        2.3 $\pm$        0.4  10$^{-17}$   &       1.3 $\pm$        0.4  10$^{-17}$   \\
   \cii\, $^2$P$_{3/2}$-$^2$P$_{1/2}$     & 157.741     &        7.5 $\pm$        0.3  10$^{-16}$   &       3.0 $\pm$        0.2  10$^{-16}$   &       2.4 $\pm$        0.5  10$^{-17}$   &        4.2 $\pm$        0.9  10$^{-17}$   &           $\le$          4  10$^{-18}$   \\
\hline
%
CO  J = 36-35     &  72.843     &        1.0 $\pm$        0.1  10$^{-16}$   &       1.5 $\pm$        0.4  10$^{-17}$   &           $\le$          5  10$^{-18}$   &            $\le$          4  10$^{-18}$   &           $\le$          1  10$^{-17}$   \\
CO  J = 33-32     &  79.360     &        2.1 $\pm$        0.2  10$^{-16}$   &           $\le$          1  10$^{-17}$   &           $\le$          9  10$^{-18}$   &            $\le$          4  10$^{-18}$   &           $\le$          1  10$^{-17}$   \\
CO  J = 29-28     &  90.163     &        3.9 $\pm$        0.2  10$^{-16}$   &       2.9 $\pm$        0.9  10$^{-17}$   &           $\le$          3  10$^{-18}$   &            $\le$          3  10$^{-18}$   &           $\le$          5  10$^{-18}$ \\  
CO  J = 18-17     & 144.784     &       1.29 $\pm$       0.03  10$^{-15}$   &       5.2 $\pm$        0.9  10$^{-17}$   &       1.3 $\pm$        0.2  10$^{-17}$   &        3.6 $\pm$        0.4  10$^{-17}$   &       1.2 $\pm$        0.2  10$^{-17}$   \\
\hline
%
p-\ho\,   6$_{15}$ - 5$_{24}$     &  78.928     &        1.0 $\pm$        0.2  10$^{-16}$   &           $\le$          1  10$^{-17}$   &           $\le$          7  10$^{-18}$   &            $\le$          3  10$^{-18}$   &           $\le$          1  10$^{-17}$   \\
p-\ho\,   3$_{22}$ - 2$_{11}$     &  89.988     &        3.3 $\pm$        0.2  10$^{-16}$   &           $\le$          1  10$^{-17}$   &           $\le$          3  10$^{-18}$   &        4.9 $\pm$        1.6  10$^{-18}$   &           $\le$          6  10$^{-18}$  \\ 
p-\ho\,   4$_{13}$ - 3$_{22}$     & 144.518     &        7.0 $\pm$        2.9  10$^{-17}$   &           $\le$          1  10$^{-17}$   &           $\le$          6  10$^{-18}$   &            $\le$          3  10$^{-18}$   &           $\le$          3  10$^{-18}$   \\
p-\ho\,   3$_{31}$ - 4$_{04}$     & 158.309     &            $\le$          1  10$^{-17}$   &           $\le$          1  10$^{-17}$   &           $\le$          6  10$^{-18}$   &            $\le$          5  10$^{-18}$   &           $\le$          4  10$^{-18}$   \\
\hline
%
o-\ho\,    8$_{18}$ - 7$_{07}$     &  63.324     &        3.2 $\pm$        1.2  10$^{-16}$   &           $\le$          4  10$^{-17}$   &           $\le$          1  10$^{-17}$   &            $\le$          4  10$^{-17}$   &           $\le$          1  10$^{-17}$   \\
o-\ho\,    7$_{07}$ - 6$_{16}$     &  71.947     &        2.9 $\pm$        0.2  10$^{-16}$   &           $\le$          1  10$^{-17}$   &           $\le$          5  10$^{-18}$   &            $\le$          4  10$^{-18}$   &           $\le$          9  10$^{-18}$   \\
o-\ho\,    4$_{23}$ - 3$_{12}$     &  78.741     &        5.6 $\pm$        0.2  10$^{-16}$   &       1.9 $\pm$        1.4  10$^{-17}$   &           $\le$          7  10$^{-18}$   &        1.2 $\pm$        0.2  10$^{-17}$   &       2.1 $\pm$        0.8  10$^{-17}$ \\  
o-\ho\,    2$_{12}$ - 1$_{01}$    & 179.527     &        6.1 $\pm$        0.2  10$^{-16}$   &       1.5 $\pm$        0.3  10$^{-17}$   &       2.6 $\pm$        0.6  10$^{-18}$   &        3.6 $\pm$        0.4  10$^{-17}$   &       1.9 $\pm$        0.6  10$^{-17}$ \\  
o-\ho\,    2$_{21}$ - 2$_{12}$    & 180.488     &        1.7 $\pm$        0.2  10$^{-16}$   &           $\le$          1  10$^{-17}$   &       1.9 $\pm$        0.6  10$^{-18}$   &        1.1 $\pm$        0.5  10$^{-17}$   &       1.0 $\pm$        0.7  10$^{-17}$   \\
\hline
%
OH    $^2 \Pi _{1/2, 1/2}$     &  79.110     &        8.9 $\pm$        0.1  10$^{-16}$   &       5.2 $\pm$        0.5  10$^{-17}$   &       8.7 $\pm$        2.4  10$^{-18}$   &        1.8 $\pm$        0.1  10$^{-17}$   &       7.7 $\pm$        3.2  10$^{-18}$   \\
OH    $^2 \Pi _{3/2, 3/2}$     &  79.180     &       1.18 $\pm$       0.01  10$^{-15}$   &       2.8 $\pm$        0.5  10$^{-17}$   &       6.0 $\pm$        2.4  10$^{-18}$   &        1.7 $\pm$        0.1  10$^{-17}$   &       1.7 $\pm$        0.3  10$^{-17}$   \\
 \hline
     \end{tabular}
\end{center}
\end{table*}

\section{Results}
\label{sect:results}

All of the observed sources show emission in a number of atomic (\oi~63.2, 145.5~\um, \cii~157.7~\um) and molecular (high-J CO, ortho and para \ho, and OH) lines (see Fig.~\ref{fig:lines_spectra} and Tab.~\ref{tab:line_fluxes}). The spatial distribution of the atomic and molecular lines is discussed in Sect.~\ref{sect:atomic} and Sect.~\ref{sect:molecular}.

   \begin{figure*}[!ht]
     \centering
     \includegraphics[width=18.cm]{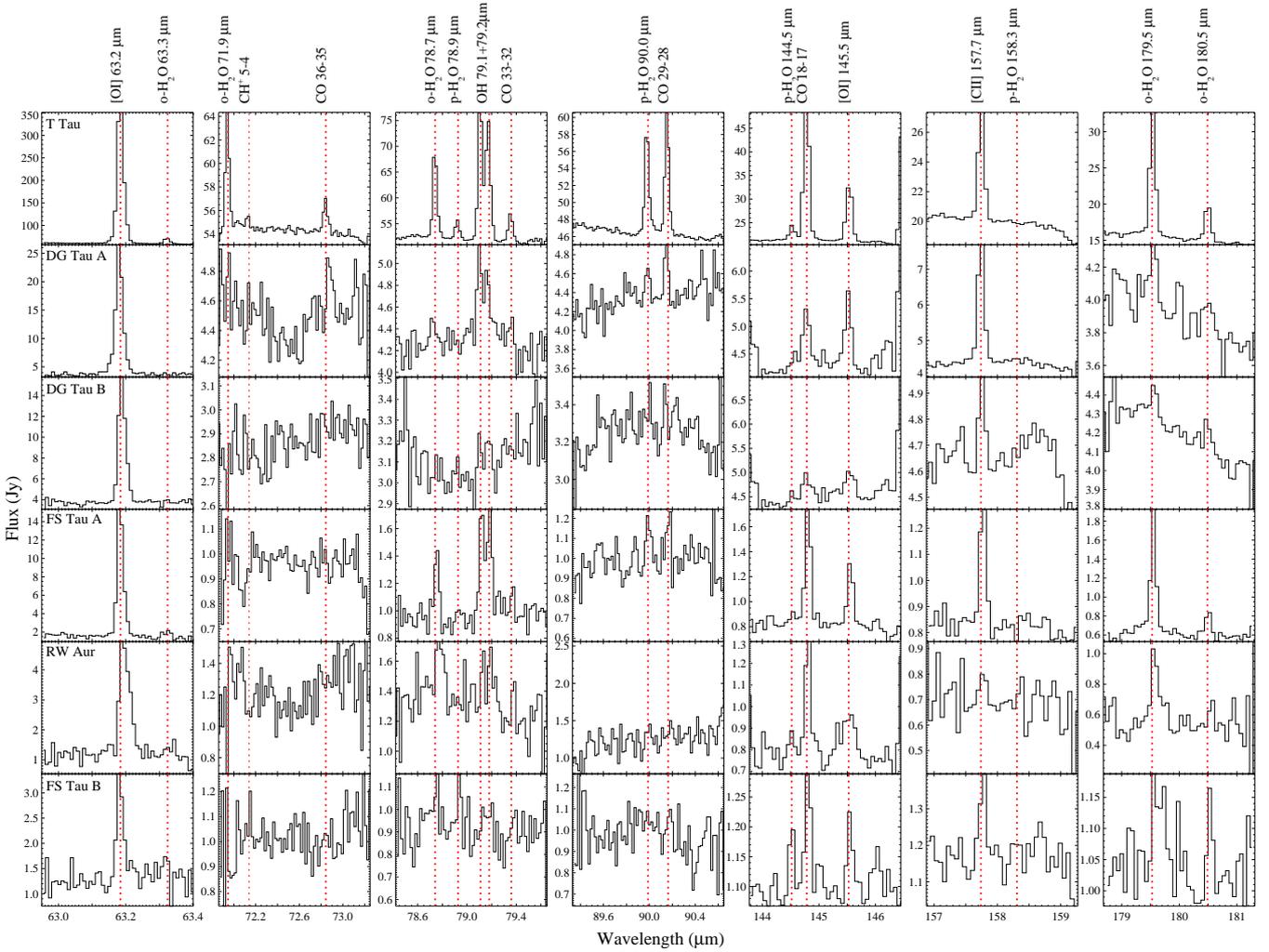}
   \caption{Spectral ranges covered by the {\it Herschel}/PACS observations for all of the sources in the sample. The wavelengths of the targeted atomic (\oi, \cii) and molecular (\ho, OH, CO) lines are shown by the vertical dotted red lines.}
   \label{fig:lines_spectra}
    \end{figure*}

    \begin{figure*}[!ht]
     \centering
     \includegraphics[width=8.cm]{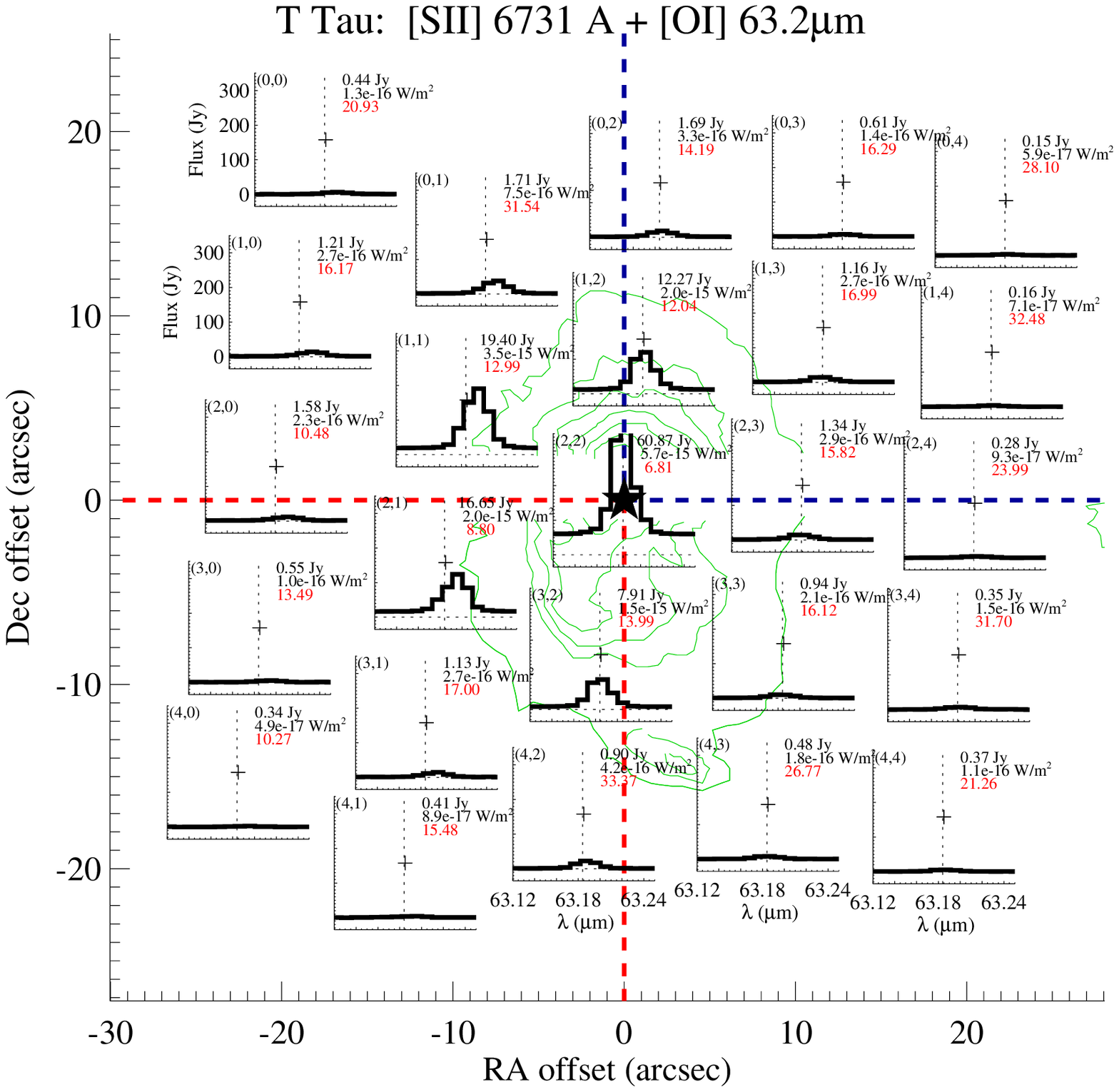}
     \includegraphics[width=8.cm]{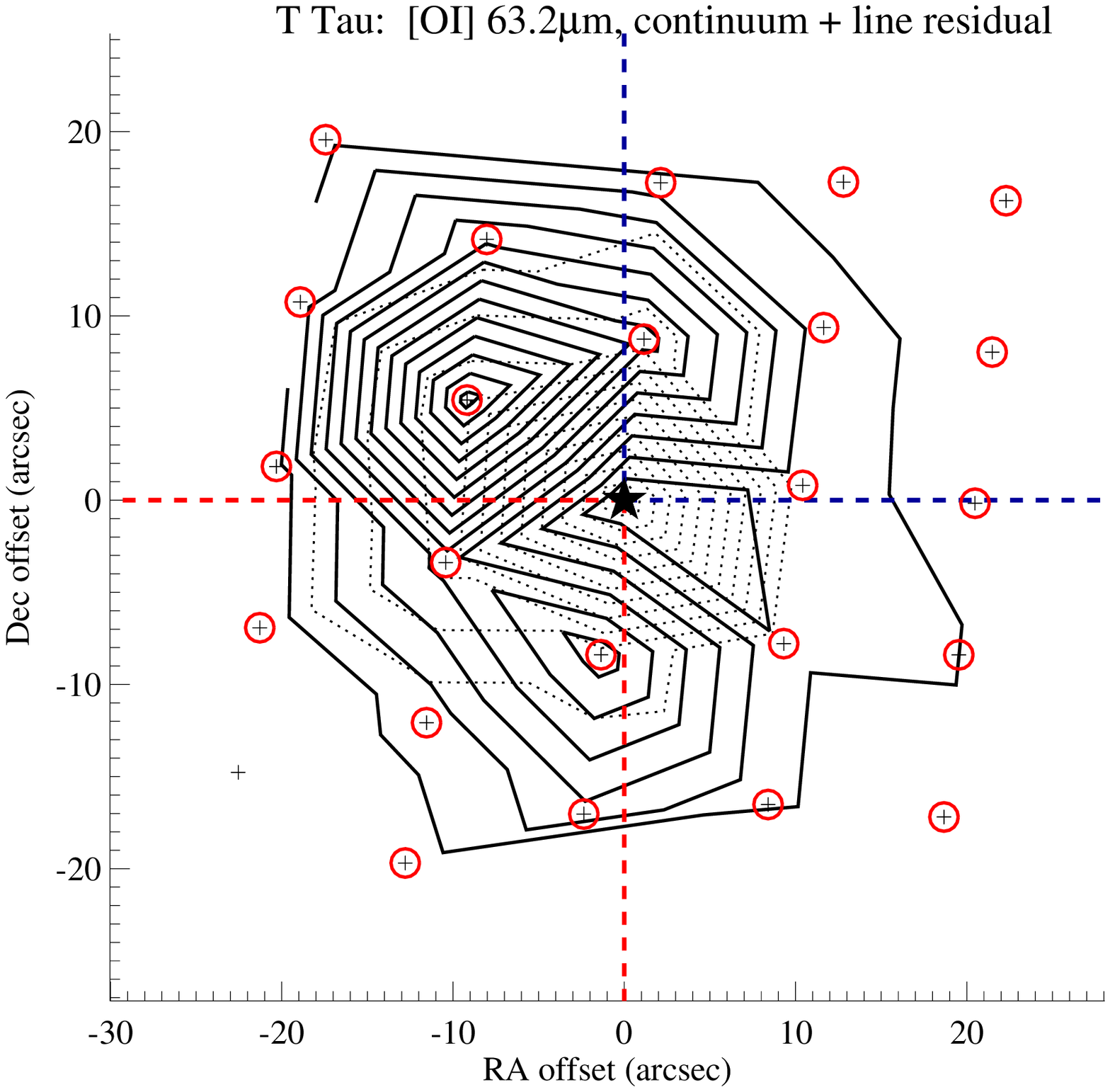}

\vspace{1.cm}

    \includegraphics[width=8.cm]{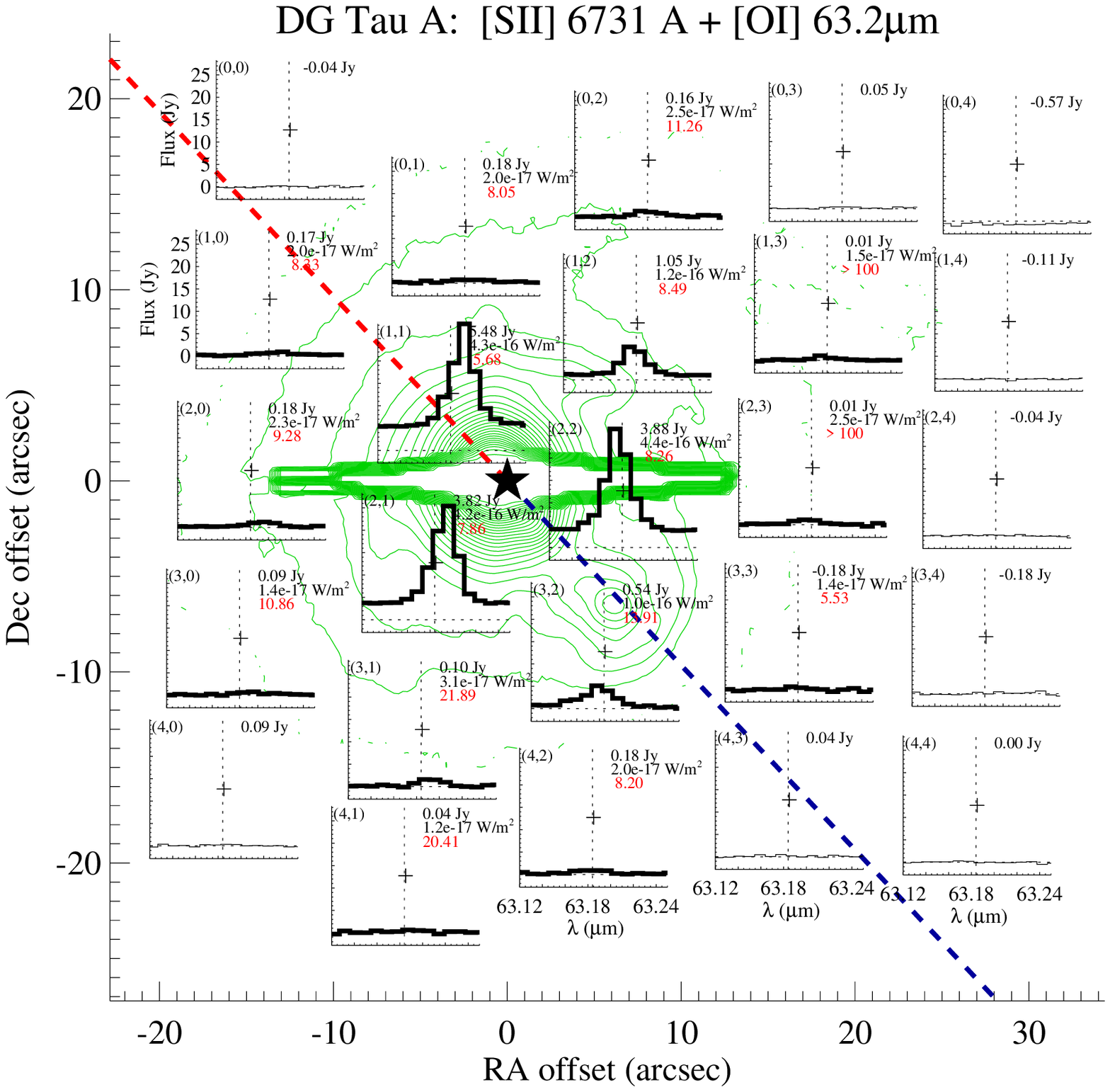}
    \includegraphics[width=8.cm]{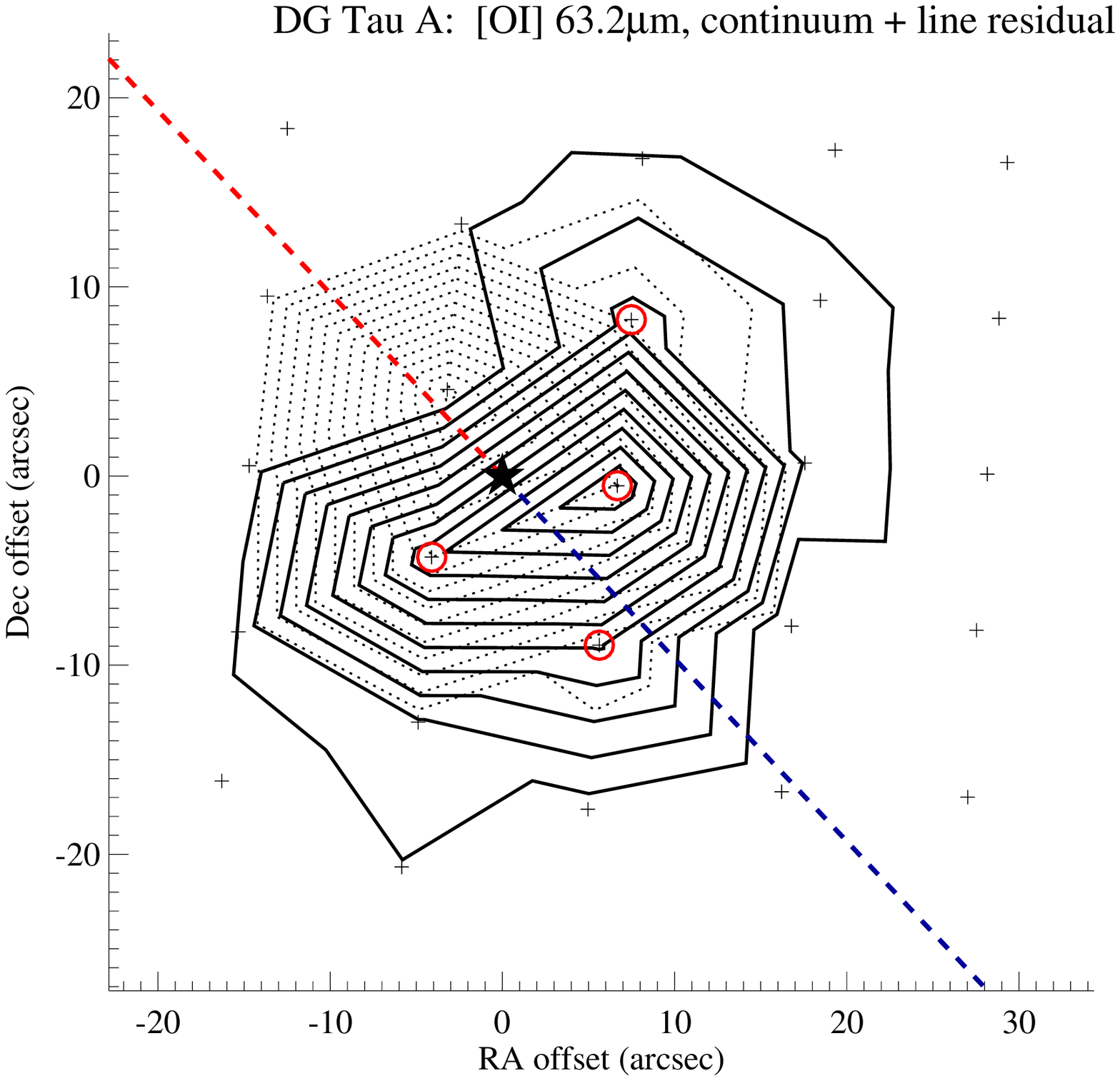}

\vspace{1.cm}

    \includegraphics[width=8.cm]{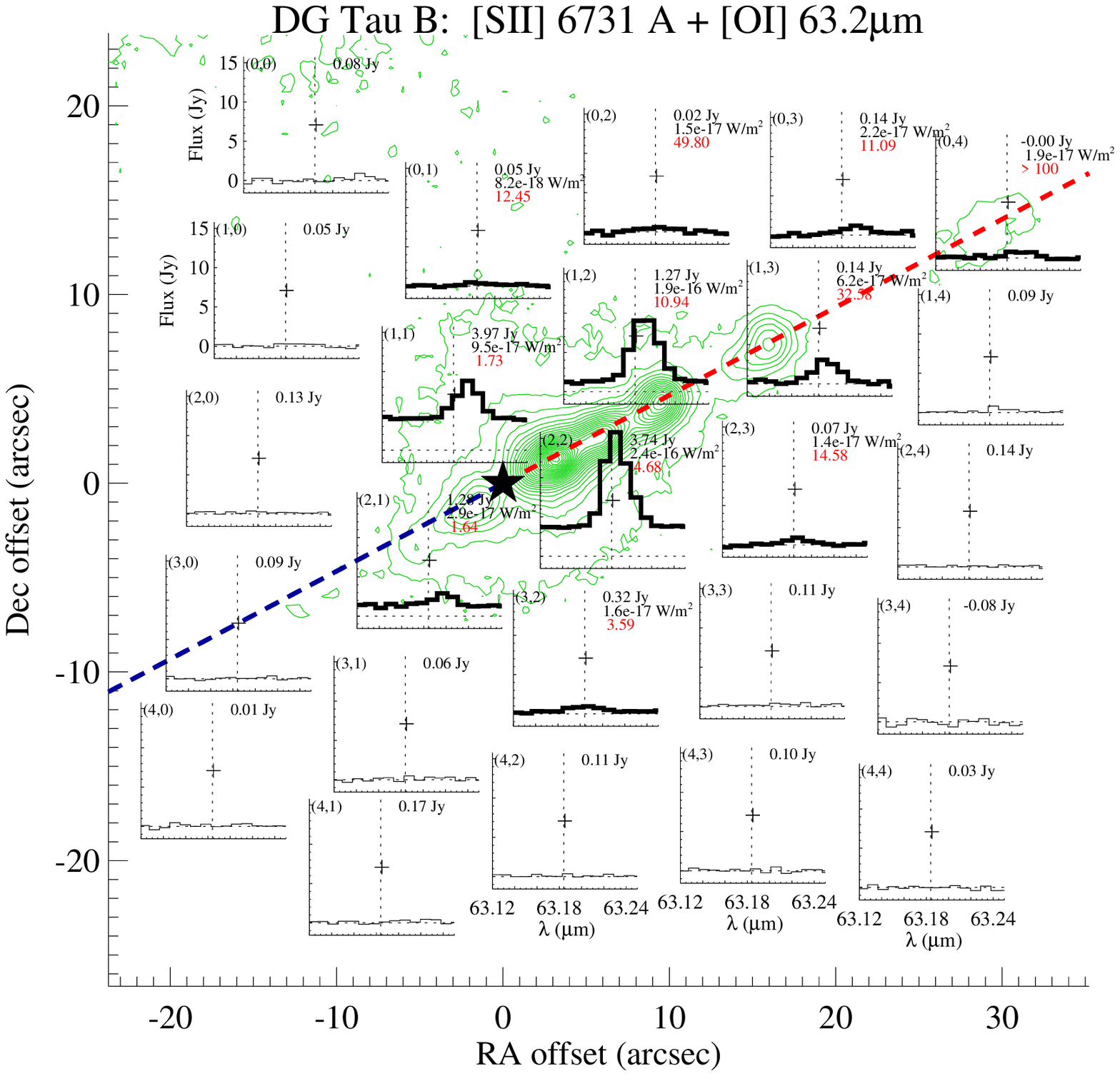}
    \includegraphics[width=8.cm]{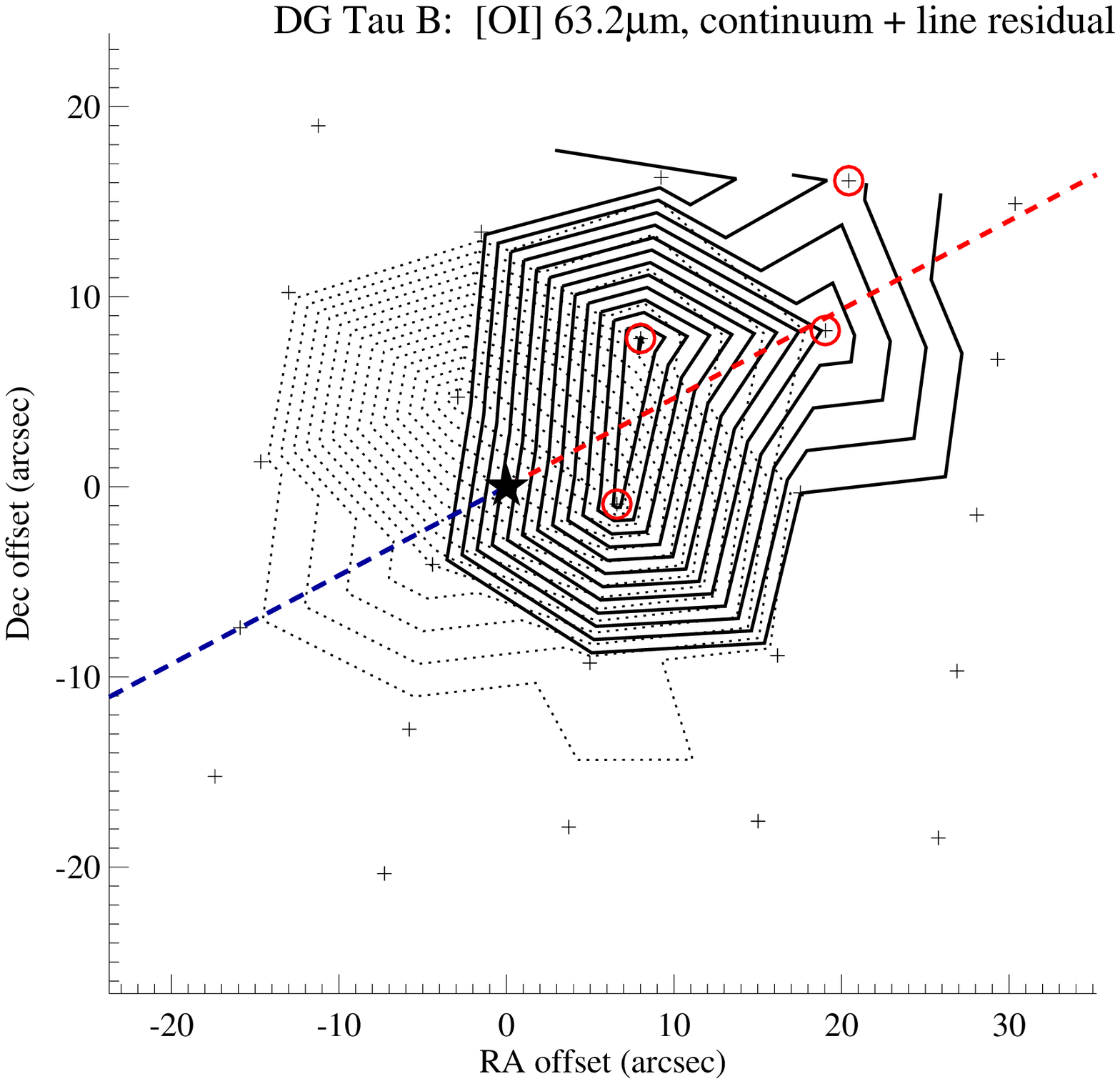}
   \end{figure*}


   \begin{figure*}[!ht]
    \centering
    \includegraphics[width=8.cm]{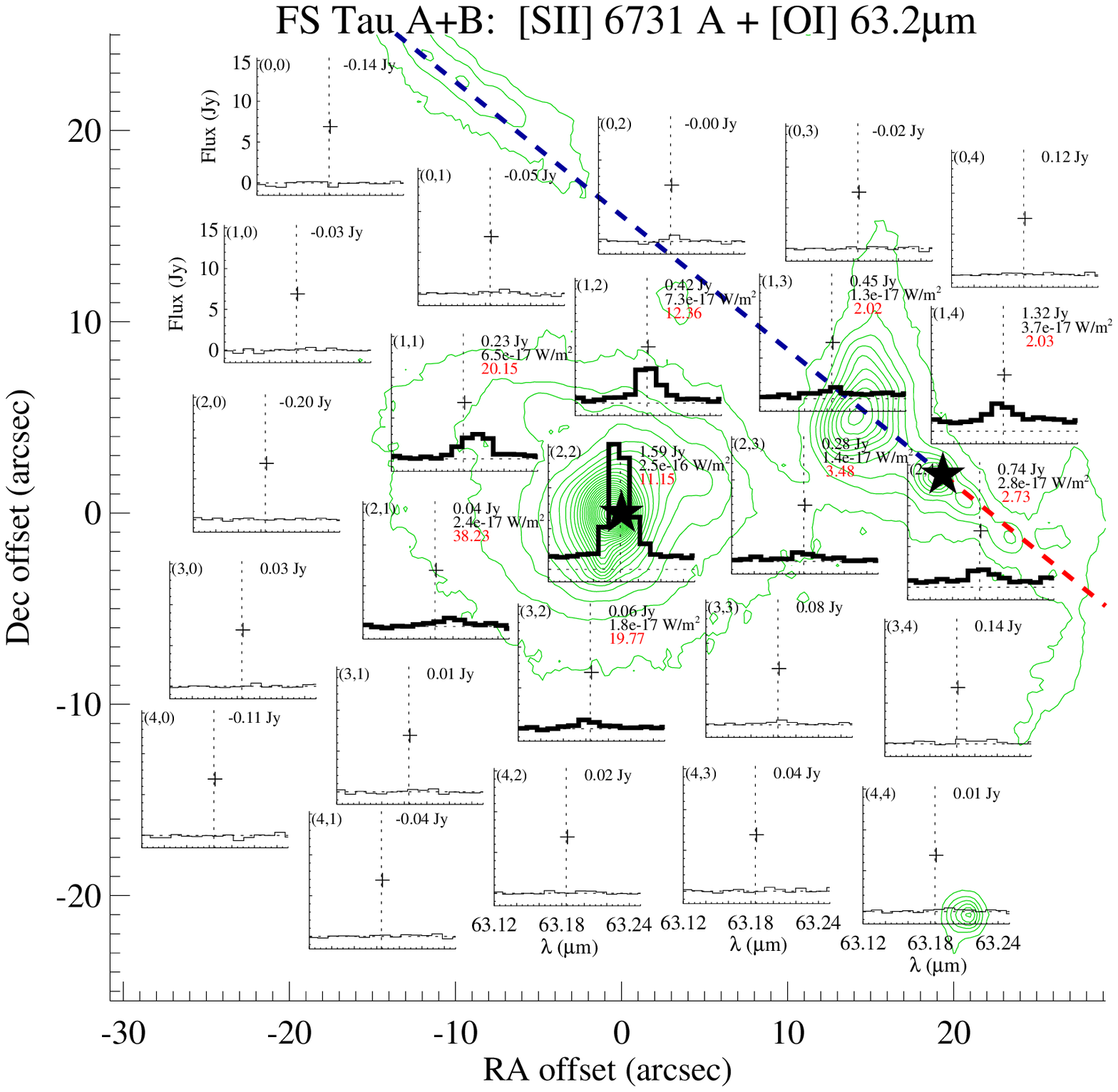}
    \includegraphics[width=8.cm]{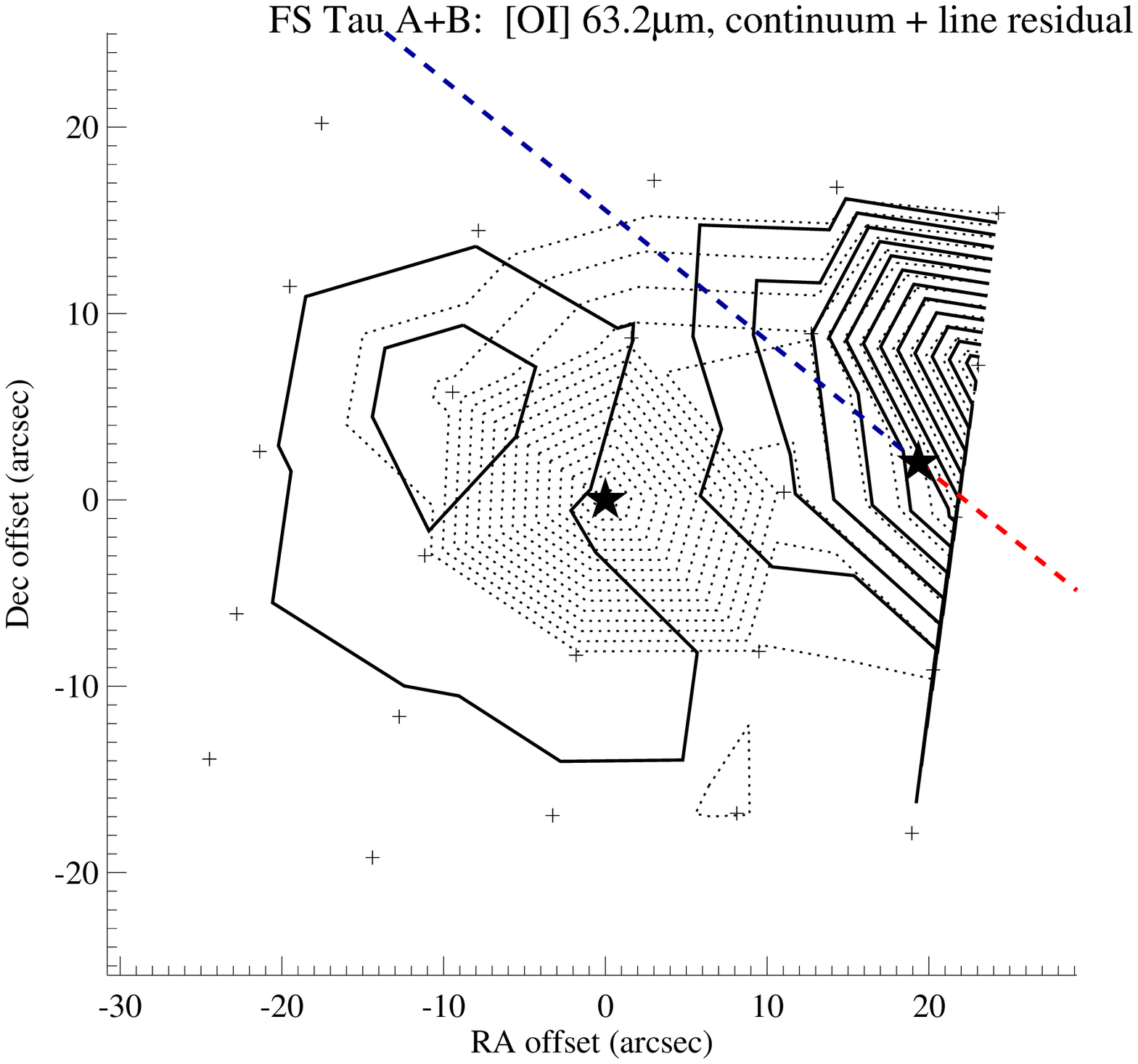}

\vspace{1.cm}

    \includegraphics[width=8.cm]{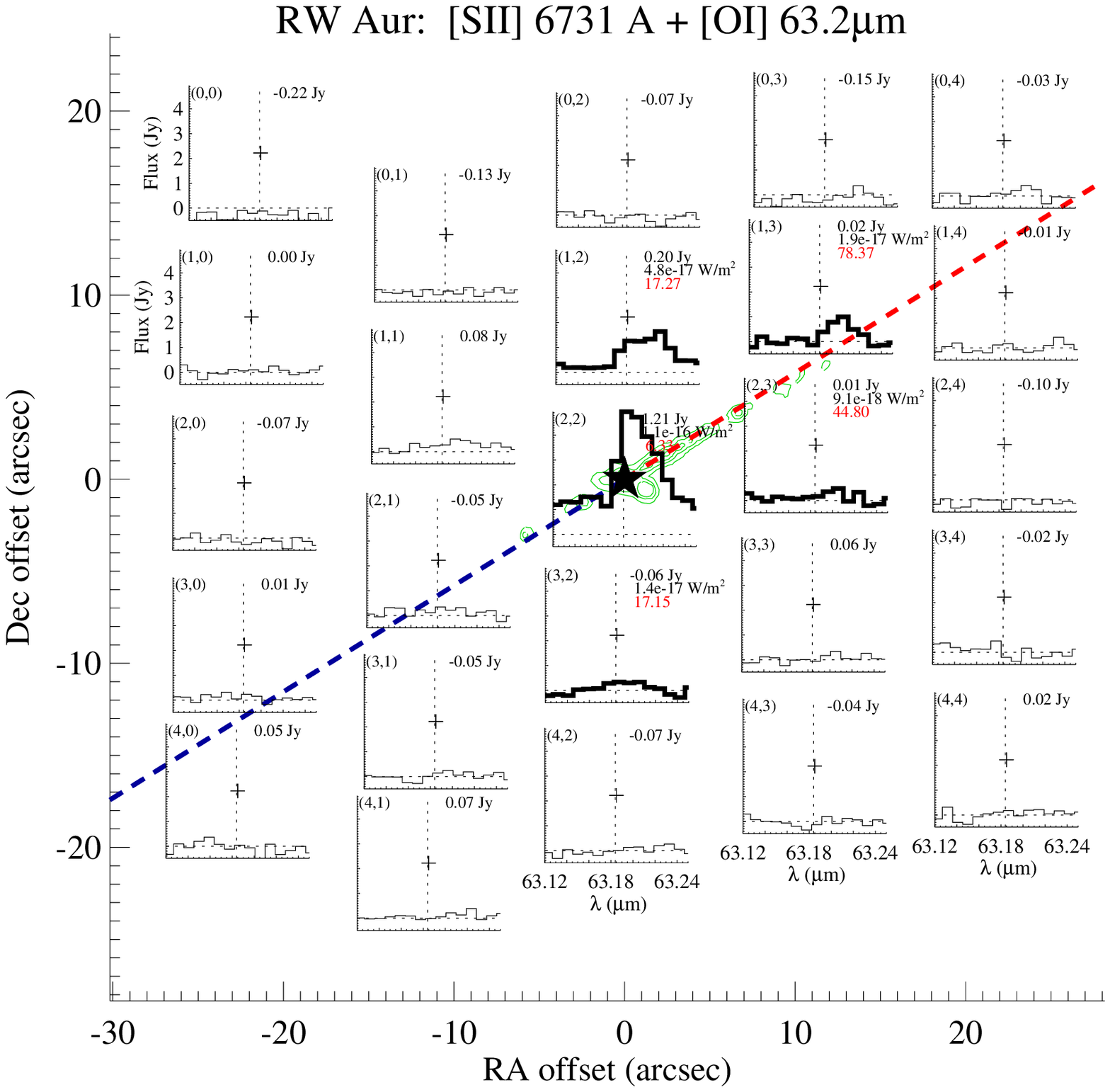}
    \includegraphics[width=8.cm]{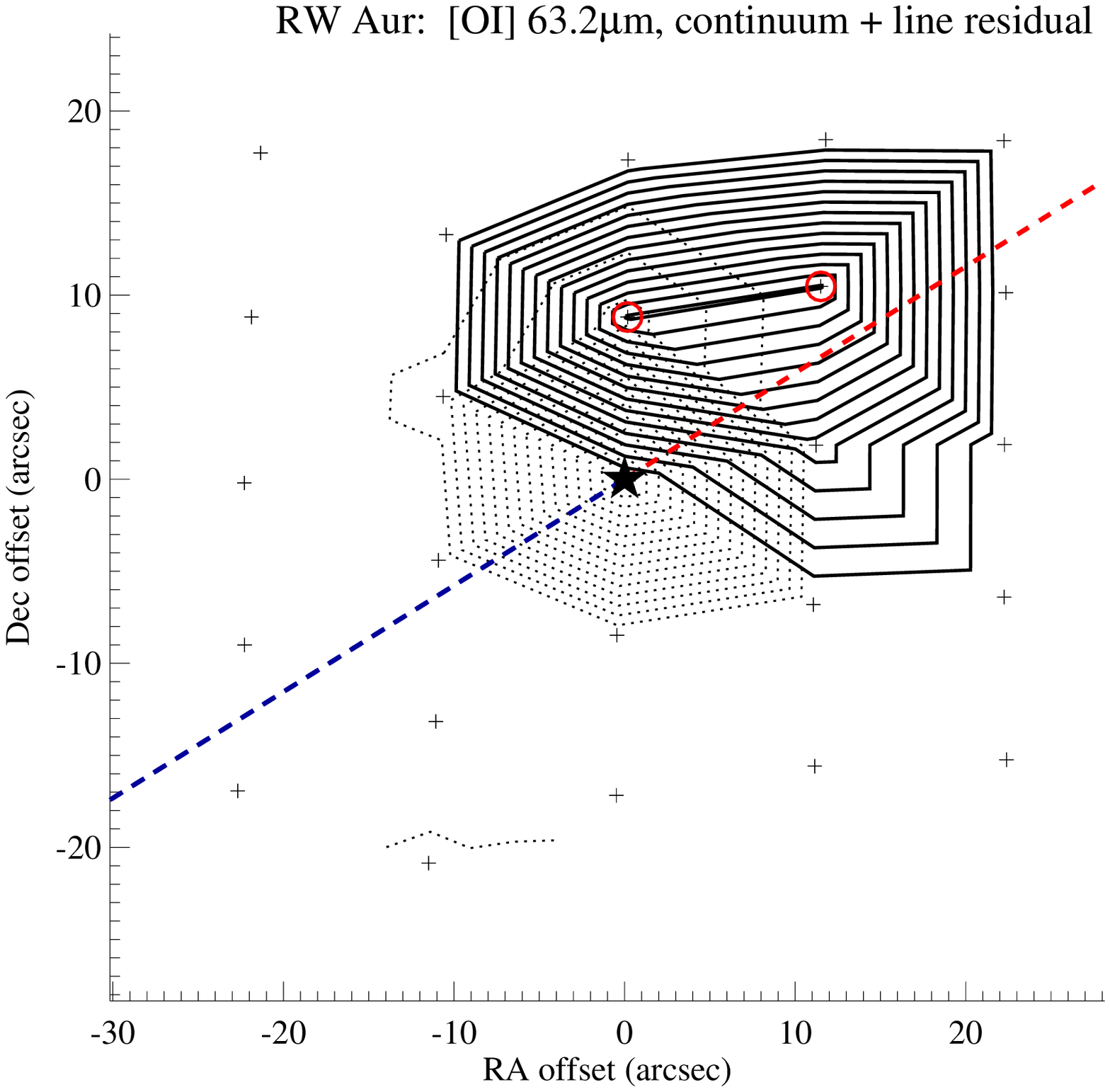}
 \caption{
{\it Left panel:} \oi~63~\um\, line spectra in the 25 spaxels of the PACS array.  Each spaxel is labelled by its (x,y) index and its position is indicated by a black cross. The RA and Declination offsets with respect to the source (in arcseconds) are indicated on the x and y axis. The black star is the source position and the blue/red dashed lines the position angle of the optical-jet blue/red lobe. In each spectrum, vertical and horizontal dotted lines show the line wavelength and the zero flux level, respectively. In those spaxels where the line is detected above the 3$\sigma$ level, the spectrum is drawn with a black thick line, and the estimated continuum level (in Jy), line flux (in \wm), and line-to-continuum ratio (in red) are indicated. Where the line is not detected only the estimated continuum level (in Jy) is given. The optical-jet emission in the \sii\lam\lam 6716, 6731 \ang\, forbidden lines is overplotted (green contours). Contours are from \citet{solf99} (T Tau), \citet{eisloffel98} (DG Tau A, DG Tau B, FS Tau A+B), and \citet{dougados00} (RW Aur).
{\it Right panel:} Contour levels of continuum (dotted lines) and line residual (solid lines) emission obtained from the analysis in Appendix \ref{app:extended_emission}. The spaxels where residual line emission is detected with a confidence level $\ge$5 are highlighted by a red circle (for RW Aur the red circles indicate residual emission above the 2$\sigma$ level). The contours indicate that the line emission is shifted and/or more extended with respect to the continuum emission along the optical jet PA. }
  \label{fig:OI_maps}
   \end{figure*}

\subsection{Atomic \oi, \cii\, emission: correlation with optical jets and millimetre outflows}
\label{sect:atomic}

 
Based on the analysis presented in Sect.~\ref{sect:ext_emission} and Appendix \ref{app:extended_emission}, all of the sources in the sample show extended emission in the \oi~63~\um\, and in the \cii~158~\um\, lines, with the exception of RW~Aur which shows faint and unresolved \cii\, emission.
Despite the limitations imposed by the low spatial resolution of the PACS data, the extended emission detected in the bright \oi~63~\um\, line is spatially correlated with the direction of the jets/outflows. 
In the following we examine the \oi~63~\um\, maps obtained for each source in the sample, and compare these with observations at optical, near-infrared and millimetre wavelengths   (see Fig.~\ref{fig:OI_maps}).\\

\noindent {\bf T Tau:}

T~Tau is a multiple system consisting of the optically visible northern component T Tau N, and the "infrared" companion T Tau S located 0$\farcs$7 to the south \citep{dyck82}. The latter is itself a close binary with a separation between the components Sa and Sb of 0$\farcs$13 \citep{kohler08}.  
T Tau N has been classified as a Class II source, while T Tau S is deeply embedded and probably a Class I source \citep{furlan06,luhman10}.
Both T Tau South and T Tau North drive jets whose forbidden optical line emission is detected up to a distance of $\sim$40\arcsec\,  along the North-South and the West-East directions \citep{solf99}.

PACS photometric observations at 70, 110 and 160 \um\, acquired within the GASPS project, and presented in Howard et al. ({\it in preparation}) show that the FIR continuum emission associated with T Tau is extended ($\ge$4\arcsec). 
In the spectroscopic observations presented in Fig.~\ref{fig:OI_maps} the unresolved multiple system is centred on the central spaxel. 

The \oi~63~\um\, line and continuum emission are maximum on the central spaxel, and are detected in all the spaxels across the PACS field of view. The line-to-continuum ratios are larger in the outer spaxels, indicating that the line emission is more extended than the continuum. 
After subtracting the on-source line emission scaled to the detected continuum level in all spaxels (see Appendix \ref{app:extended_emission} for details), we detected residual line emission above the 5$\sigma$ confidence level in most of the outer spaxels up to a distance of $\sim$28\arcsec.
The residual emission shows two peaks located $\sim$10\arcsec\, to the north-east (F$_{residual} \sim$ 1.7~10$^{-15}$ \wm) and $\sim$8\arcsec\, to the south (F$_{residual} \sim$ 8.2~10$^{-16}$ \wm) with respect to the source.
At the PACS resolution it was impossible to distinguish the emission associated with the two jets detected at optical wavelengths but only possible to identify the presence of extended line emission. 
Note that the bulk of the \oi\,63 \um\, line emission originates from a region of $\le$11\arcsec\, around the source but extends up to $\sim$30\arcsec (i.e. up to the edge of the PACS grid) similarly to what is seen in the optical forbidden lines \citep{solf99}.
\\
  
\noindent {\bf DG Tau A:}

DG Tau A is a strongly accreting Class II source \citep[e.g., ][]{hartigan95,luhman10} associated with a jet. The DG Tau A jet was first detected at optical wavelengths by \citet{mundt83}.
High angular resolution studies at optical and NIR wavelengths showed a bright collimated blue-shifted lobe moving with radial velocities up to $\sim$350 \kms, whilst only faint emission was detected on the red-shifted side \citep[e.g., ][]{dougados00, pyo03}.
Spectroscopic observations taken with the Hubble Space Telescope and presented by \citet{bacciotti00} and \citet{maurri12} indicated that the bulk of the emission in the \oi\,6300 \ang, and in the other optical forbidden lines (\sii\, and \azii\, lines) comes from the first 1$\farcs$3 of the blueshifted jet lobe.
Beyond 1$\farcs$3, two strong bow-shocks are detected. One is at $\sim$3$\arcsec$-4$\farcs$5 from the source and the other is at $\sim$9-10\arcsec\, from the source. Both working surfaces are moving with a proper motion of $\sim$0.3$\arcsec$/year.

Fig.~\ref{fig:OI_maps} (left panel) shows that in our PACS observations the source is not centred in the central spaxel but lies 6$\farcs$7 to the east with respect to it. Thus, both the line and the continuum emission are detected in a number of spaxels around the source position.
However, whilsty the continuum emission peaks at the position of spaxel (1,1), the \oi~63~\um\, line peaks at the position of spaxel (2,2), i.e. $\sim$6$\farcs$7 to the west with respect to the source. 
This indicates that the line emission is offset with respect to the continuum, and is possibly more extended.
The  right panel of Fig.~\ref{fig:OI_maps} shows the residual line emission offset with respect to the continuum emission, and displaced along the direction of the blue lobe that is detected at optical wavelengths.
The residual line emission reaches its maximum $\sim$7\arcsec\, from the source (F$_{residual} \sim$1.5 10$^{-16}$ \wm) and extends $\sim$11\arcsec.
\\

\noindent {\bf DG Tau B:}

DG Tau B has been classifies as a Class I source by \citet{luhman10,rebull10} and is associated with the bright, asymmetric HH 159 jet \citep{mundt83, eisloffel98}.
The red lobe consists of a chain of bright knots detected in the optical forbidden lines (\azii, \oi, \sii) that extends $\sim$55\arcsec\, from the source. The fainter blue lobe is only detected up to $\sim$10\arcsec\, from the source in the same lines.
Observations at millimetre wavelengths in the CO low-J lines confirm the asymmetric structure showing a slow, wide-angle red-shifted outflow which is displaced along the axis of the optical jet, while blue-shifted CO emission is faint and confined on-source \citep{mitchell94, mitchell97}. 
The non-detection of the blue lobe at millimetre wavelengths suggests that the observed asymmetry is not due to obscuration of the blue lobe but to different physical conditions in the two lobes, where the bright red lobe is two times slower and less ionized but denser, and more collimated than the faint blue lobe \citep{podio11}. 

The \oi~63~\um\, emission map in  the left panel of Fig.~\ref{fig:OI_maps} further confirms this asymmetric structure.  The peak of the line emission (in spaxel (2,2)) is not coincident with the peak of the continuum emission (at spaxel (1,1)), and the line emission is displaced $\sim$34\arcsec\,  (i.e. up to the edge of the PACS grid) along the direction of the optical-jet/mm-outflow red lobe. 
Residual emission above the 5$\sigma$ level was detected in four spaxels along the red-lobe jet PA, decreasing from 1.5 10$^{-16}$ \wm\,   at $\sim$7\arcsec\, down to 2 10$^{-17}$ \wm\, at 26\arcsec\, from the source (see Fig.~\ref{fig:OI_maps}, right panel). 
An order of magnitude decrease in flux between the first few arcseconds of the jet and the emission at 20\arcsec-30\arcsec\, from source was also seen in the optical forbidden lines \citep{podio11}.  
Therefore, the PACS maps suggest that the \oi~63~\um\, line is tracing a warm outflow component, intermediate between the hot, fast and collimated jet traced by the optical forbidden lines and the cold and slow wide-angle outflow traced by millimetre CO emission. Higher angular and spectral resolution observations are required to fully analyse the spatio-kinematical structure of this warm component. 
\\

\vspace{1.cm}
\noindent {\bf FS Tau A, B:}

FS Tau A is a Class II close binary system (separation $\sim$0$\farcs$23 by \citealt{white01}, class by \citealt{furlan06, luhman10}), associated with a bright reflection nebulosity but showing no clear evidence of an associated jet \citep{krist98,eisloffel98}.
FS Tau B is an embedded not optically visible Class I source \citep{luhman10,rebull10} located $\sim$20\arcsec\, to the west. This source is driving a parsec scale collimated outflow \citep{mundt84,eisloffel98}.

The \oi~63~\um\, map in the left panel of Fig.~\ref{fig:OI_maps} shows two continuum peaks, at the position of spaxel (2,2) and (1,4) corresponding to the position of FS Tau A and FS Tau B, respectively.
Line emission was detected in a number of spaxels located along the PA of the optical jet associated with FS Tau B (PA$\sim$55\degr, \citealt{mundt84}) and around FS Tau A.
However, because of the limited spatial resolution of PACS, it is not possible to separate the line and continuum emission associated with the two sources and to apply the analysis presented in Appendix \ref{app:extended_emission}.
Thus,  the total line and continuum emission from the FS Tau system (Aa+Ab+B) is considered throughout the paper.   
\\

\vspace{1.cm}
\noindent {\bf RW Aur:}

RW Aur is a binary system ($\sim$1$\farcs$4 separation) and has been classified as a Class II source \citep{white01, furlan06}.
RW Aur A is associated with an asymmetric jet whose bright red lobe is detected in the optical \sii, \azii, and \oi\, forbidden lines up to $\sim$15\arcsec\, from the source \citep[e.g., ][]{hirth94,hirth97,dougados00,melnikov09}.

The binary system is not resolved with PACS. The emission map at 63~\um\, in Fig.~\ref{fig:OI_maps} shows that
both the continuum and line emission peak on the central spaxel but the \oi~63~\um\, line is detected also in a few outer spaxels located along the direction of the optical jet red lobe (PA$\sim$120\degr, \citealt{hirth94,hirth97,dougados00}). This suggests the line is tracing the warm FIR counterpart of the hot optical jet.   
After subtracting on-source emission there is no residual emission above 5$\sigma$. However, residual emission above 2$\sigma$ is observed along the red-shifted jet direction at $\sim$9\arcsec\, and $\sim$15\arcsec\, from the source (F$_{residual} \sim$1.8 10$^{-17}$ \wm). 
\\

Since all of the sources in our sample show extended \oi~63~\um\, emission spatially correlated with the direction of the optical jets we also expect the line profile to be blue- or red- shifted in agreement with what is observed at optical wavelengths. However, the low spectral resolution offered by PACS ($\sim$88 \kms\, at 63~\um) is further limited by the fact that extended emission can broaden, and shift, the line spectral profile (see PACS manual).
In the case of T Tau, DG Tau A, and FS Tau A+B this effect clearly dominates, as we can see that the line peak shifts from red to blue velocities moving across the PACS field of view (see Fig.~\ref{fig:OI_maps}).
However, in the case of DG Tau B and RW Aur the \oi~63~\um\, spectral profile is much larger than the instrumental one (FWHM up to 200 \kms) and consistently shows a red-shifted peak velocity in all the spaxels displaced along the direction of the optical jet red lobe.
The peak velocity estimated from those spaxels showing residual line emission $\ge$5$\sigma$ is of $\sim$45 \kms\, and $\sim$95 \kms\, for DG Tau B and RW Aur, in agreement with values estimated from optical forbidden lines \citep{podio11,melnikov09}.\\
   
Even if for all of the sources in the sample extended emission is detected in the atomic \oi\, and \cii\, lines suggesting a jet/outflow origin, the emission from the disk may dominate on-source. For this reason we compare observed atomic lines with predictions from both disk and shock models in Sect.~\ref{sect:discussion}.

\subsection{Molecular \ho, CO, and OH lines: unresolved emission and high excitation
lines}
\label{sect:molecular}

   \begin{figure}[!ht]
     \centering
     \includegraphics[width=8.8cm]{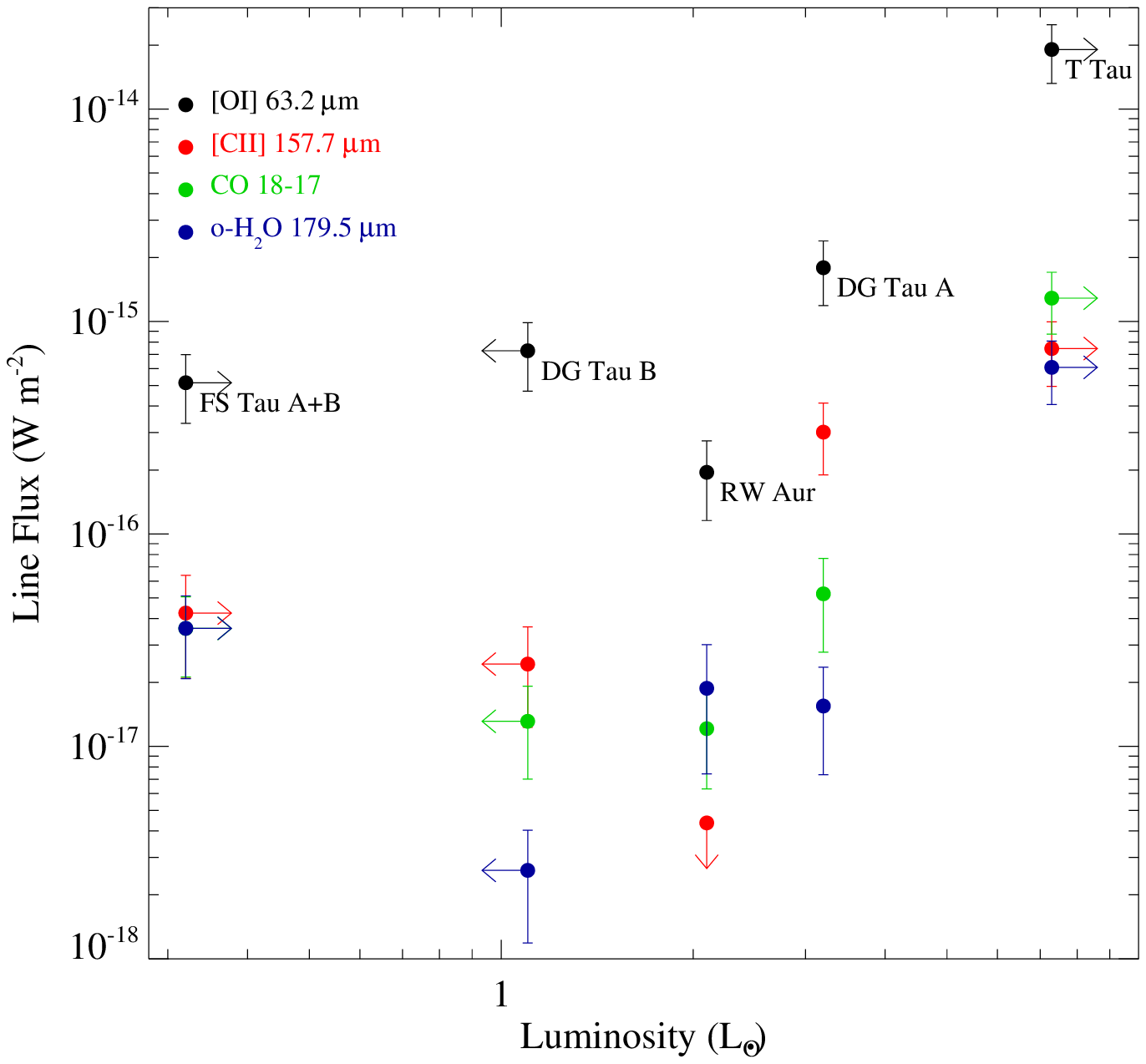}
     \includegraphics[width=8.8cm]{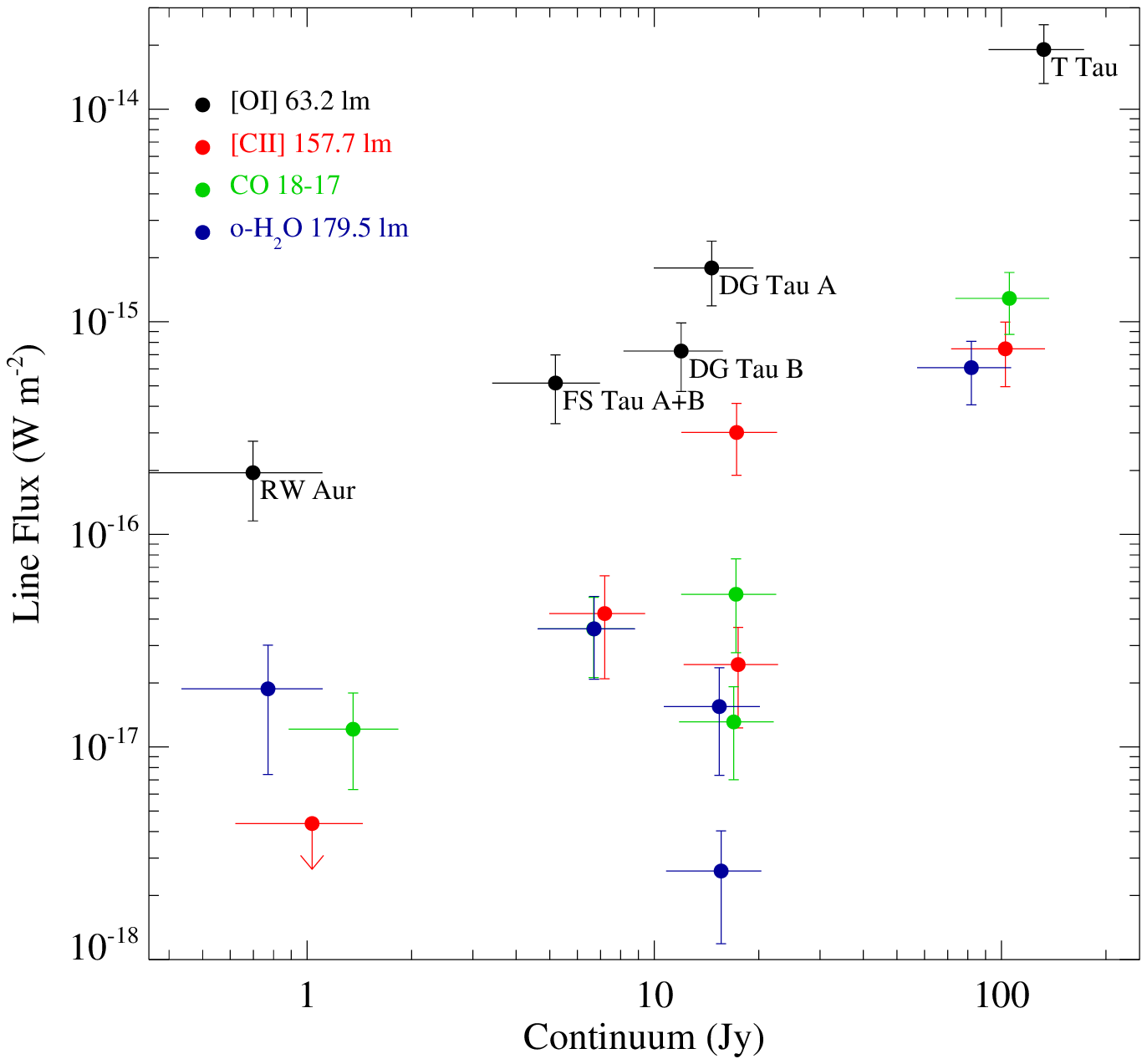}
  \caption{The fluxes of the \oi~63~\um, the \cii~158~\um,  the CO J=18-17, and the  o-\ho~179.5~\um\, lines are plotted versus the source luminosity ({\it left panel}) and their adjacent continuum ({\it right panel}) for all the sources in our sample (black, red, green, and blue dots). For FS Tau A+B and T Tau N+S the plotted source luminosity is a lower limit, since L$_{*}$ of the Class I components FS Tau B and T Tau S is unknown. The upper limit on the stellar luminosity for DG Tau B corresponds to its bolometric luminosity as estimated by \citet{kruger11}.}
   \label{fig:line_cont_lum}
    \end{figure}


In our brightest target, T Tau, we have detected five ortho-\ho\, and four para-\ho\, lines, including lines from highly excited levels, such as the o-\ho\,8$_{18}$ - 7$_{07}$ line (E$_{up} \sim$1070 K).
In the other sources, only a few of the lower excitation \ho\, lines were detected, up to the o-\ho\,4$_{23}$ - 3$_{12}$ line (E$_{up} \sim$432 K). T Tau and DG Tau also show emission from high excitation CO levels up to CO J=36-35 line (E$_{up} \sim$3668 K). Only the lowest CO transition covered by our observations, i.e. the CO J=18-17 line (E$_{up} \sim$944 K), is detected from our other sources (see Table \ref{tab:line_fluxes}). 

We have also detected the J=5-4 CH$^{+}$ emission line at 72.14 \um\, from T Tau. The other CH$^{+}$ transitions falling in the spectral range covered by our observations (i.e. the J=4-3 and J=2-1 lines at 90.01 and 179.60 \um, respectively) are blended with the p-\ho\, and o-\ho\, lines at 89.988 and 179.527 \um. These are the brightest transitions of para and ortho water in all of our sources. Because of the low spectral resolution of PACS and that the water lines are an order of magnitude brighter than the detected CH$^{+}$ emission line ($\sim$4 10$^{-17}$ \wm), we cannot deconvolve the lines and recover the flux of the faint CH$^{+}$ lines.
The origin of the spectrally and spatially unresolved CH$^{+}$ line is unclear. This line could be excited in the disk \citep{thi11} and/or in the directly irradiated outflow cavity walls \citep{bruderer10}.

On the contrary to that found for the atomic \oi\, and \cii\, lines, the emission in the molecular lines is both spectrally and spatially unresolved with PACS.
By applying the analysis in Appendix \ref{app:extended_emission} we found that, even when line emission is detected out of the central spaxel, the line-to-continuum ratio is constant across the PACS grid and there is no residual line emission above 5$\sigma$ after the subtraction of on-source line and continuum. Hence, the line and continuum emission detected in the outer spaxels are due to the spectroscopic PSF and the line and continuum emitting regions are unresolved. This implies that the atomic and molecular lines have different spatial distributions and origins. While the extended atomic emission is clearly associated with the optical jets, the compact molecular lines may be excited in the disk, in the UV-heated outflow cavities, and/or in the first 9\arcsec\, ($\sim$1300 AU) along the jet. To understand which of these components dominates the line emission, we investigate possible scenarios in the following section.

\section{Discussion}
\label{sect:discussion}

  \begin{figure*}[!ht]
     \centering
     \includegraphics[width=8.8cm]{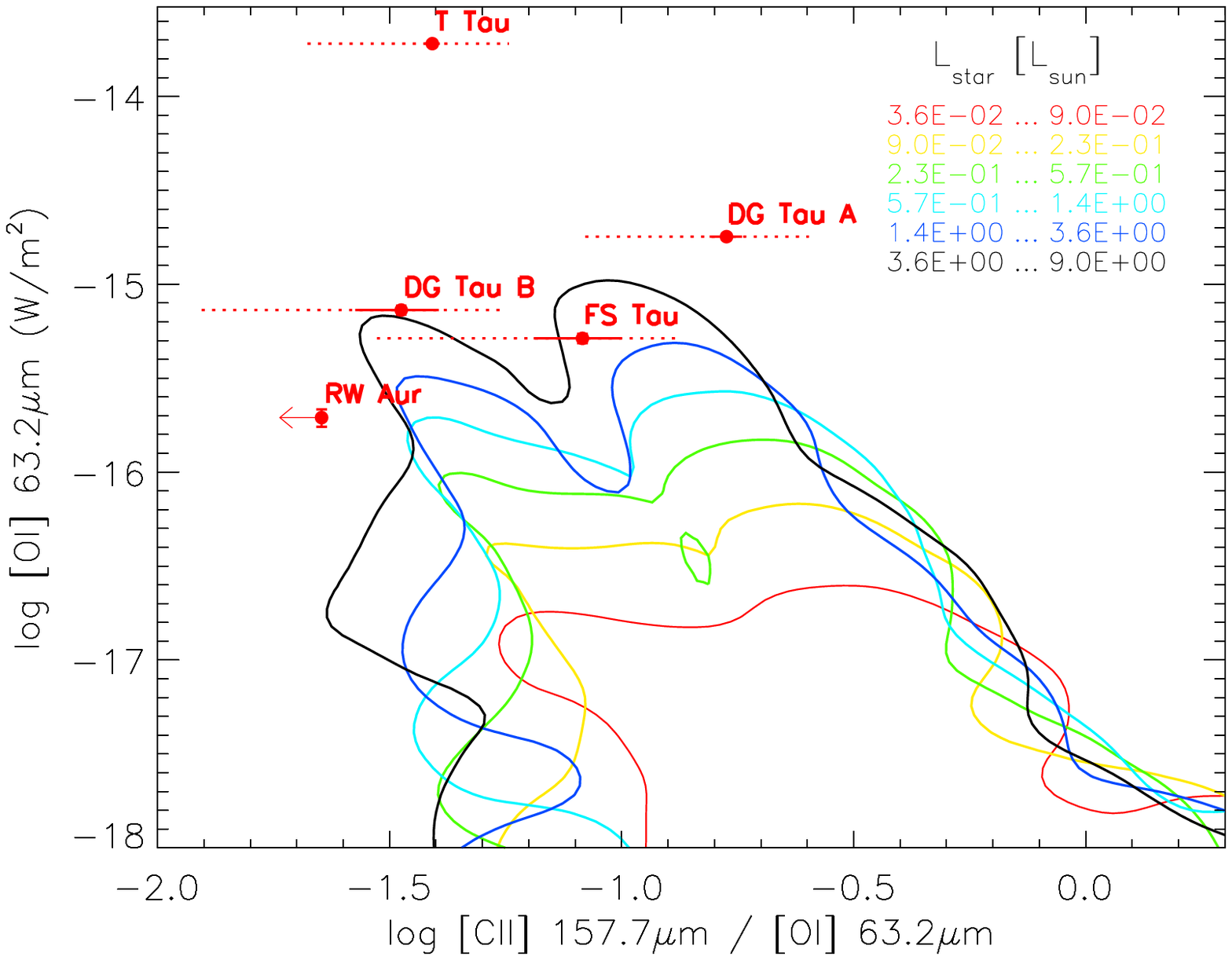}\\

     \vspace{0.2cm}
     \includegraphics[width=8.8cm]{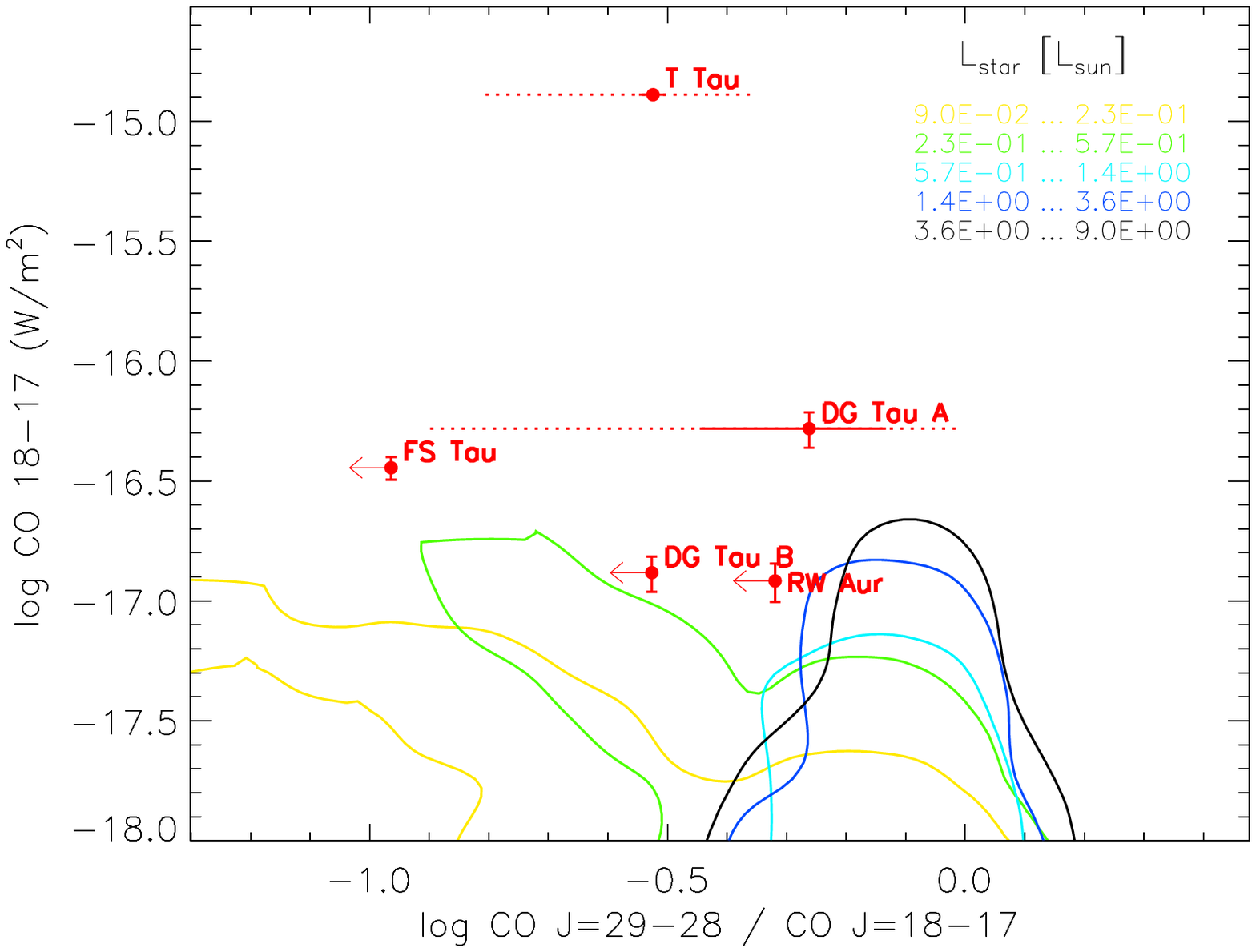}
     \includegraphics[width=8.8cm]{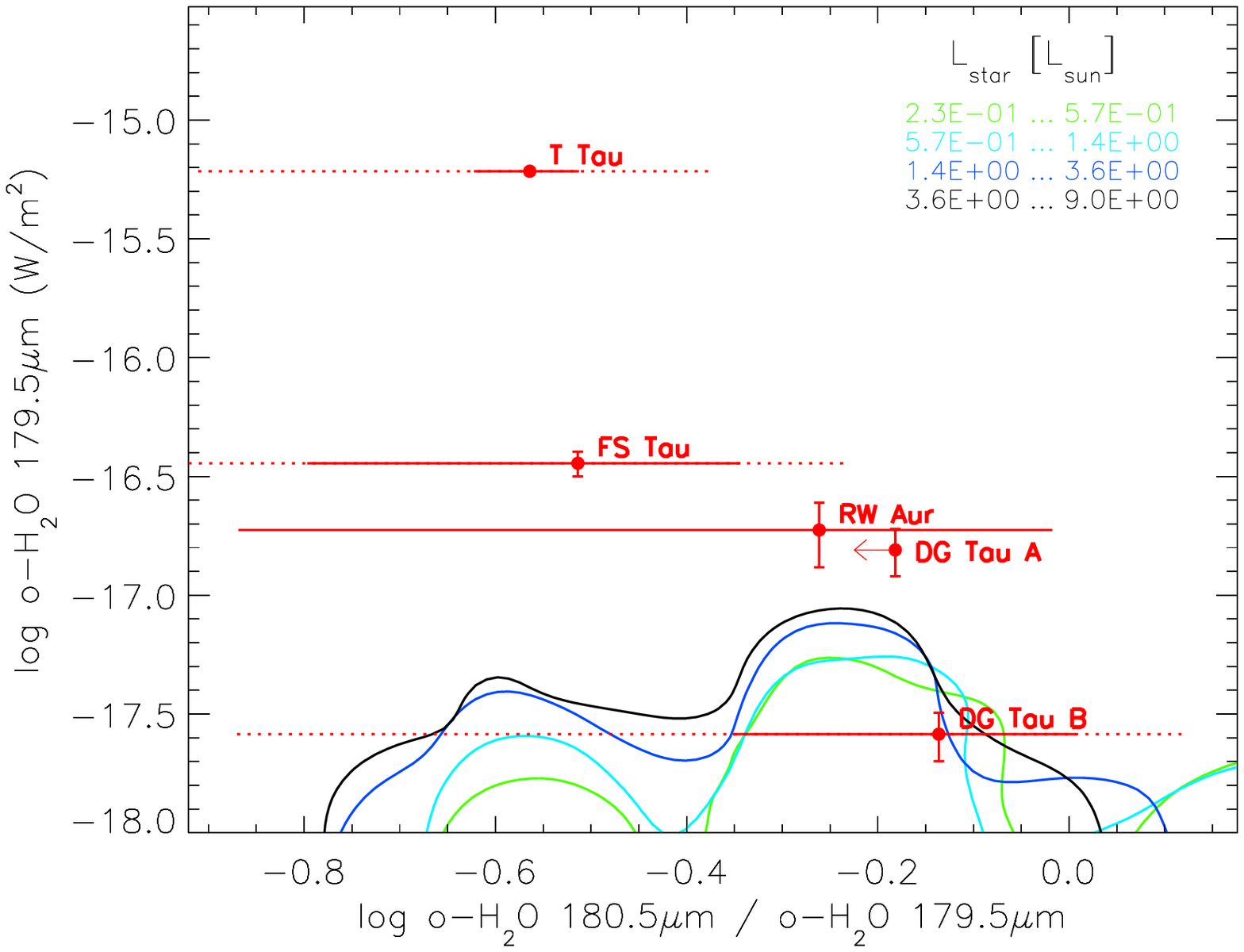}
 \caption{Observed \oi, \cii, CO, and \ho\, line fluxes and ratios (red dots) are compared with predictions from a subsample of disk models from the DENT grid \citep{woitke10,kamp11}. Contours encircle 85\% of the DENT disks for star luminosity values from 10$^{-2}$ \lsol\, to 9 \lsol\, (red, yellow, green, cyan, blue, and black crosses). The total errors on the observed values (dotted red lines) are obtained summing the error due to the line signal-to-noise (solid red lines) and the 30\% error affecting PACS flux calibration.  The subsample of disk models is obtained by selecting M$\le$2.5 \msol, T$_{eff}$ $\le$ 5500 K, L$\le$9 \lsol, R$_{in}$=R$_{subl}$, and dust-to-gas ratio = 0.01.}
   \label{fig:disk_models}
    \end{figure*}

Fig.~\ref{fig:line_cont_lum} shows the fluxes of the brightest observed lines (\oi~63.2~\um, \cii~157.7~\um, \ho~179.5~\um, CO J=18-17) versus the source luminosity (left panel) and the continuum flux adjacent to the considered line (right panel).
The plots show no clear correlation, except for the \oi~63~\um\, line. 
Howard et al. ({\it in preparation}) examine a large sample of Taurus sources, including optical-jet sources (i.e. sources with jet signatures, such as forbidden emission lines which are extended and/or show blue-shifted profiles), and sources showing no evidence of outflowing activity.
Most of the Class II jet-sources in Taurus are associated with micro-jets extending only up to a few arcseconds away from source in the typical optical tracers. These sources show unresolved \oi~63~\um\, emission at PACS resolution, making it difficult to disentangle jet and disk emission.
However, Howard et al. ({\it in preparation}) find that optical-jet sources show excess \oi~63~\um\, emission with respect to the tight F~(\oi~63~\um) - Continuum~(63~\um) correlation which is found for sources with no evidence of outflowing activity.  By using this correlation and the measured continuum fluxes we find that the emission from the disk may account for 3$\%$ to 15$\%$ of the observed \oi~63~\um\, line flux.  This suggests that in jet sources emission from the jet/outflow dominates over possible disk emission in the \oi~63~\um\, line.

To verify this hypothesis and constrain the origin of the unresolved molecular emission in the following sections we compare the atomic and molecular line fluxes, and their ratios, with disk models and shock models predictions (Sect.~\ref{sect:disk_models} and Sect.~\ref{sect:shock_models}, respectively).  
The line ratios are computed by using the line fluxes summarised in Table \ref{tab:line_fluxes}. 
Unfortunately, an analysis of the variation of line ratios with distance from source is not possible from several reasons, i.e. (i) in some observations the source is not centred on the central spaxel; (ii) because of the spectroscopic PSF the emission in the outer spaxels is strongly contaminated by on-source emission overlapping to local extended emission; (iii) the source position on the PACS field of view may be different when observing at different wavelengths.  

\subsection{Emission from disks}
\label{sect:disk_models}


To understand how much the disk can contribute to the extended \oi\, and \cii\, atomic lines and to the unresolved molecular emission, we compare observed line fluxes and ratios with predictions from the DENT grid of disk models \citep{woitke10,pinte10,kamp11}. 
The DENT grid consists of 300~000 disk models spanning a large range of parameters defining the source (mass, M$_{*}$, temperature, T$_{eff}$, luminosity, L$_{*}$, UV excess) and the disk (disk gas mass, gas-to-dust ratio, inner and outer disk radius, R$_{in}$, R$_{out}$, surface density, flaring, dust grain size distribution, dust settling, disk inclination) properties.
For each model, the dust temperatures, gas temperatures and chemical structure are computed to produce a large set of observables such as the SED and selected FIR and submillimetre line fluxes.
We consider a sub-sample of models in the grid corresponding to typical low-mass young stellar objects (YSO) and T Tauri star properties (M$_{*}$ $\le$ 2.5 \msol, T$_{eff}$ $\le$ 5500 K, L$_{*}$ $\le$ 9 \lsol). We also assume that the inner disk radius, R$_{in}$, is located at the position of the dust sublimation radius, R$_{subl}$ \citep{pinte08} and that the dust-to-gas ratio is 0.01, as expected in the case of young primordial disks.

Fig.~\ref{fig:disk_models} indicates that the selected disk models cover a large range of line ratios but cannot reproduce the absolute fluxes of the observed bright lines. In particular, we observe \oi~63~\um\, fluxes up to 1.9 10$^{-14}$ \wm\, and 1.8 10$^{-15}$ \wm\, for T Tau and DG Tau A, which cannot be produced by the disk even for high star luminosity. DG Tau B and FS Tau A+B show lower \oi\, fluxes ($\sim$5-7 10$^{-16}$ \wm) but, if we consider that their luminosity is lower than 1 \lsol, disk models cannot account for the observed flux.
The emission in the fine structure lines from disk models is similar to that predicted by PDR models, with the fundamental difference that it originates over a wider range of density/temperature. The \oi\, lines are optically thick and close to LTE (the critical densities for the \oi~145~\um\, and \oi~63~\um\, lines are 6~10$^4$, 5~10$^5$ cm$^{-3}$ respectively: \citealt{kamp10}). 

The discrepancy between observed fluxes and predictions from disk models is even more evident for the molecular lines: while observed CO 18-17 and \ho~179.5~\um\, line fluxes are of 10$^{-17}$-10$^{-15}$ \wm\, and  3~10$^{-18}$-6~10$^{-16}$ \wm, the predicted fluxes are always lower than a few  10$^{-17}$ \wm\, for the CO 18-17 line and 10$^{-17}$ \wm\, for the \ho~179.5~\um\, line.  
According to the disk models these lines originate at the disk surface and the line excitation temperature determines the radial extent of the emitting area. For solar-type stars these water lines are optically thick and not in LTE \citep{aresu12}. The high-J CO lines are generally in LTE due to their low critical densities, and they can be optically thick depending on the details of the model.
Higher atomic and molecular line fluxes can be obtained by assuming a lower dust-to-gas ratio (dust-to-gas = 0.001) and very large inner radii (e.g., R$_{in} >$10 R$_{subl}$). These values, however, are typical of more evolved disks but are not appropriate to describe the young disks observed around Class I and II sources.
Since some of the observed sources are strong X-rays emitters \citep{gudel07a}, we also checked the effect of X-rays on the considered emission lines in the work of \citet{aresu11} and \citet{meijerink12,aresu12}. It can be shown that for the considered subsample of disk models the line fluxes are not significantly enhanced by X-rays (less than a factor of 2).

\subsection{Emission from shocks}
\label{sect:shock_models}


\begin{figure}[!ht]
    \centering
    \includegraphics[width=8.8cm]{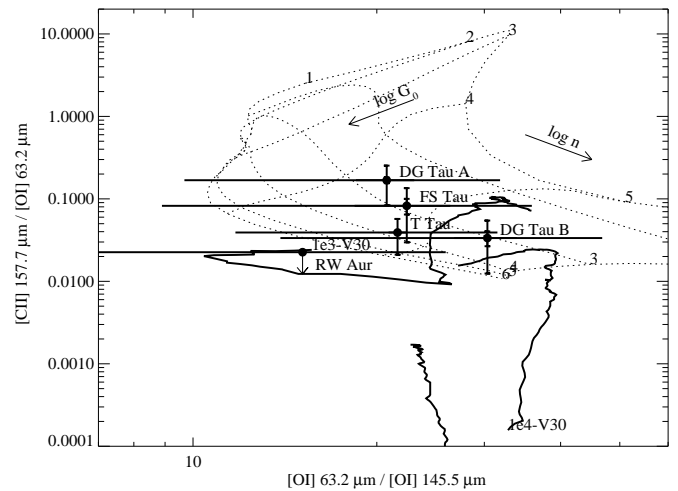}
 \caption{Observed atomic line ratios are compared with predictions from PDR (dotted lines, \citealt{kaufman99}) and fast J-type shock (solid lines, \citealt{hollenbach89}) models. For the PDR models, the labels on the broken curves indicate the gas density ($n$) and the intensity of the FUV field ($G_0$), respectively. For the shock models, the pre-shock density is marked at the lowest shock--velocity point (V$_{\rm shock}$=30 \kms), and the shock velocity increases along the full curves, up to V$_{\rm shock}$=150 \kms.}
  \label{fig:atomic_ratios}
   \end{figure}


The strong spatial correlation between the \oi\, 63 \um\, line and the optical jets and the large observed fluxes suggest that the emission in the atomic lines originates from the shocks occurring along the jet and/or from the UV-heated gas in the outflow cavity walls, rather than in the circumstellar disk. Thus, we compare atomic line fluxes and ratios with predictions from shock models and PDR models.

Fig. \ref{fig:atomic_ratios} shows the observed atomic line ratios \oi63\um/145\um, hereafter \oi63/145, and \cii158\um/\oi63\um, hereafter \cii/\oi, and the predictions by PDR models \citep{kaufman99} and fast J-type shock models, in which H$_2$ is fully dissociated and the gas is partially ionized in a radiative precursor \citep{hollenbach89}.
The lower--velocity (V$_{shock}$$\sim$10--40 \kms) C- and J-type shock models of \citet{flower10}, which do not incorporate a radiative precursor, are absent from this plot, as they predict \cii~158~\um\, emission from 2 to 8 orders of magnitude lower than the \oi~63~\um\, emission with increasing pre-shock density.

Fig. \ref{fig:atomic_ratios} indicates that both PDR models with densities $>10^4$~\cmc\, and FUV field G$_{0}>10^3$ and  dissociative J-type shocks with low pre-shock density (n$\sim$10$^3$-10$^{4}$ cm$^{-3}$, V$_{shock}$ = 30-150 \kms) are able to reproduce the \oi63/145 and the \cii/\oi\, line ratios for most of the sources in our sample. The exceptions are RW Aur, which cannot be reproduced by PDR models because \cii/\oi~$\le$0.02, and DG Tau A, which cannot be reproduced by the fast J-shock models because of its large \cii/\oi\, line ratio ($\sim$0.17).

The \oi~63~\um\, maps presented in Fig.~\ref{fig:OI_maps} show that the \oi\, line emission originates from a region extending up to 30\arcsec\, from the source. Such extended emission could not be produced by UV-illuminated outflow cavities. 
Thus a shock origin likely dominates the \oi\, line emission.
However, part of the observed emission may originate from the illuminated outflow cavities, which would also explain \cii/\oi\, line ratios higher than what is predicted by shock models.
Note that the only object showing no evidence of surrounding cloud material or outflow cavities is RW Aur, which is also the only source in our sample with no detected \cii~158~\um\, emission.


 \begin{figure*}[!ht]
    \centering
    \includegraphics[width=8.8cm]{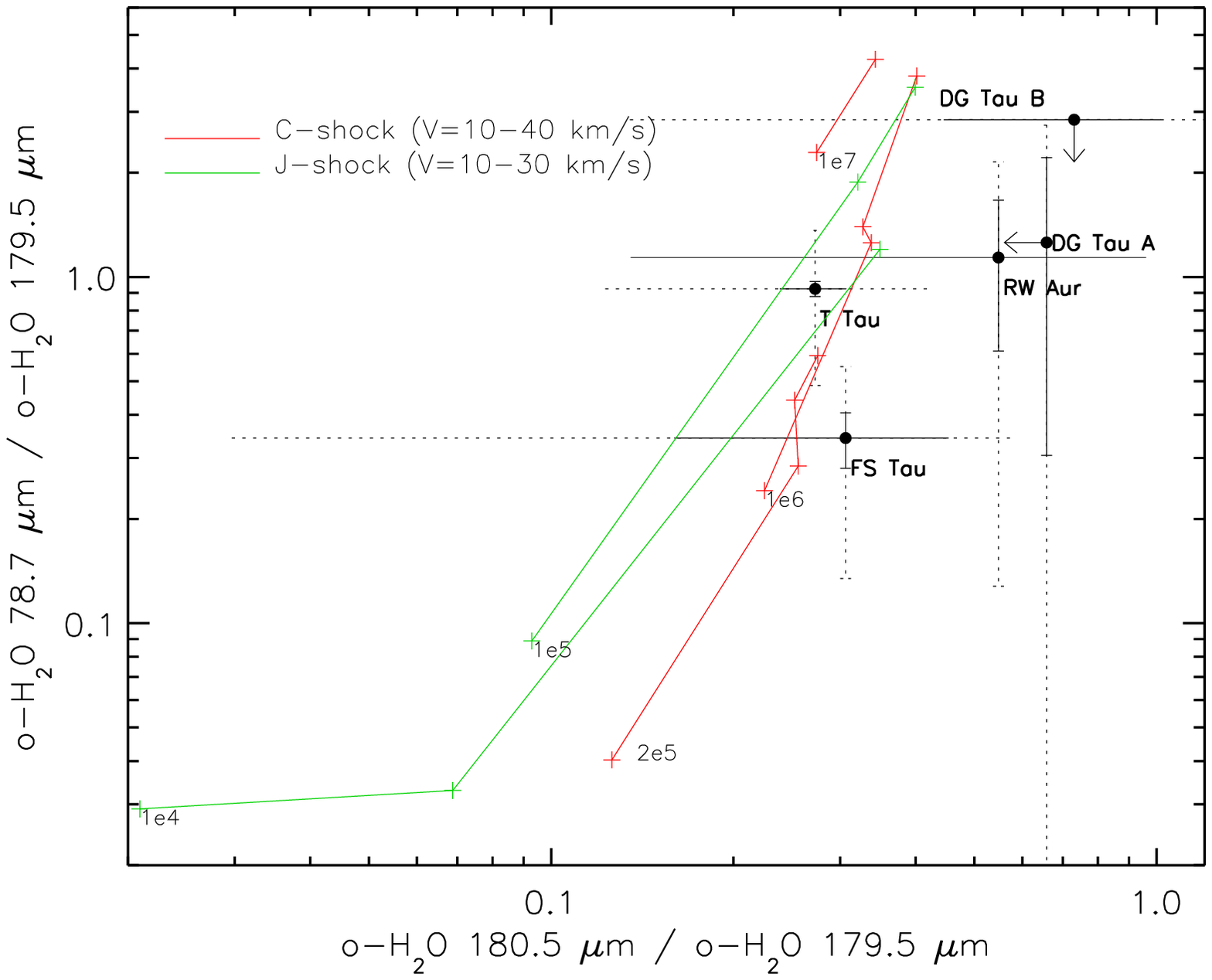}
    \includegraphics[width=8.8cm]{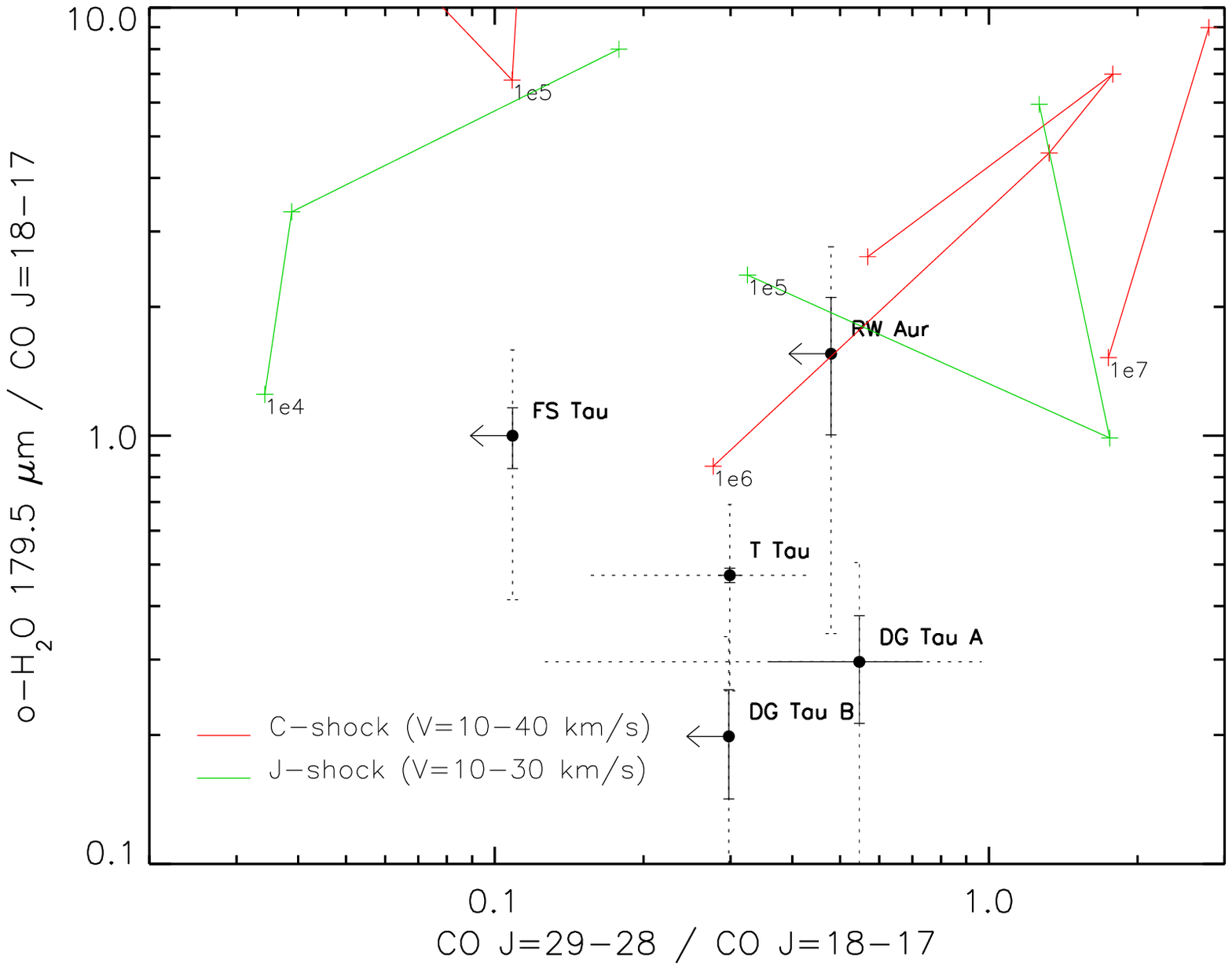}
 \caption{Observed \ho\, and CO line ratios (black dots) are compared with the predictions of slow C-type ({\it red lines}) and J-type ({\it green lines}) shock models \citep{flower10}. The solid lines indicate the errors due to the line signal-to-noise, while the dashed lines indicate the total error, obtained by adding the 30\% error in the PACS flux calibration.  The crosses along the red and green curves correspond to increasing shock velocity, from 10 \kms\, to 40 \kms\, (C-type shocks) or from 10 \kms\, to 30 \kms\, (J-type shocks). At the lowest velocity (V$_{\rm shock}$=10 \kms), the pre-shock density is given.}
  \label{fig:h2o_ratios}
   \end{figure*}

  \begin{figure*}[!ht]
    \centering
    \includegraphics[width=8.8cm]{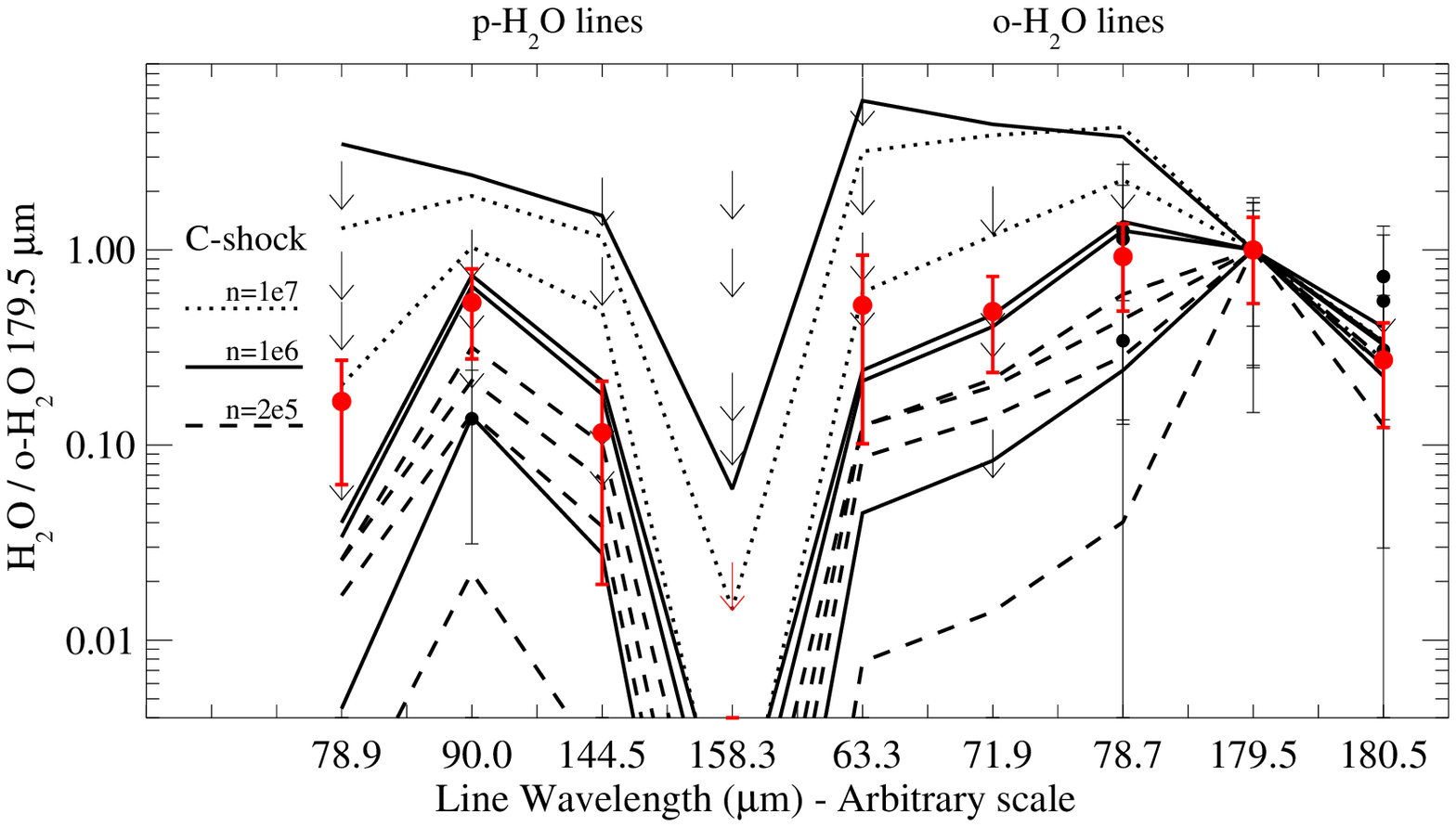}
    \includegraphics[width=8.8cm]{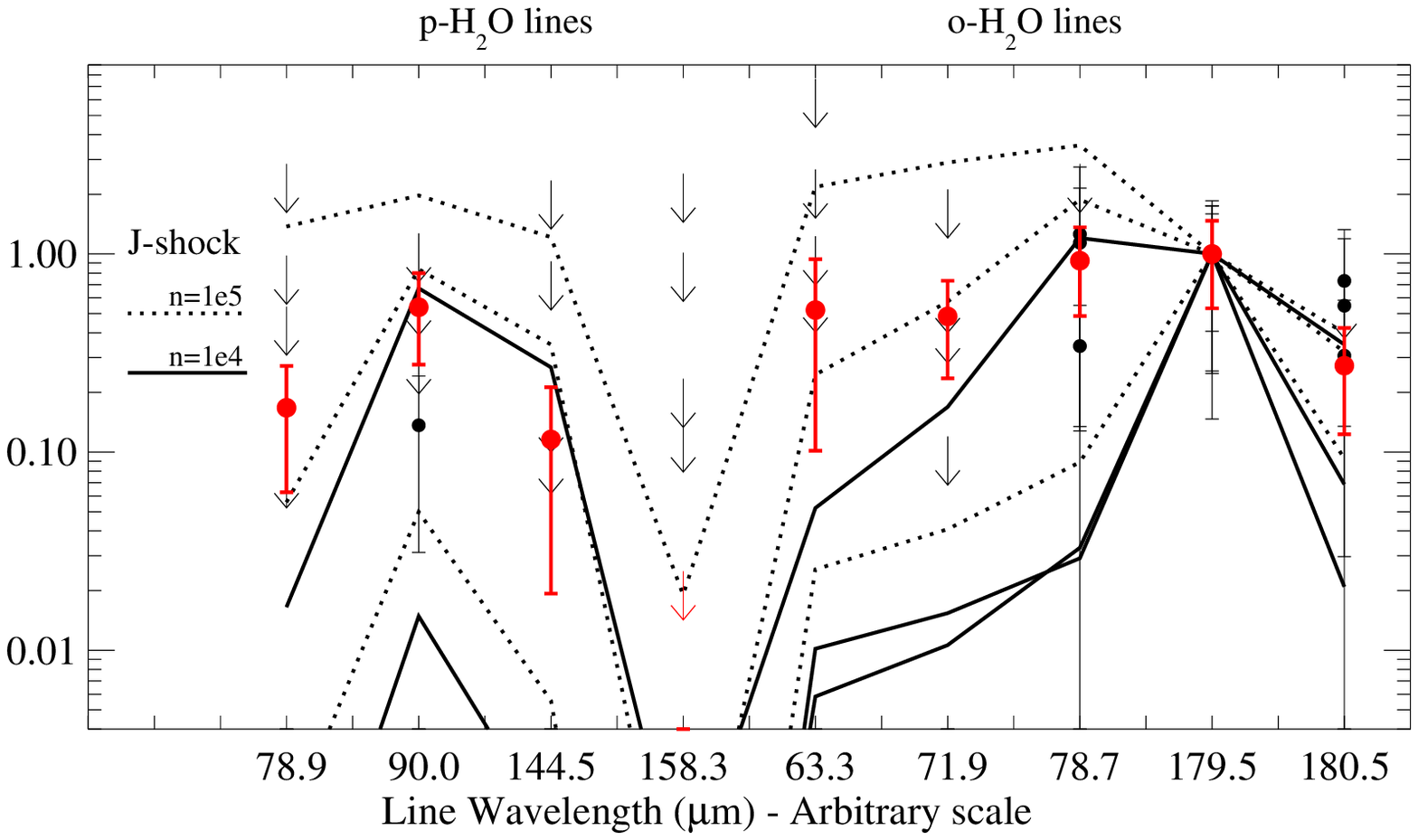}
 \caption{Observed \ho\, line ratios for T Tau (red points/arrows) are compared with the predictions of C-type ({\it left panel}) and J-type ({\it right panel}) shock models \citep{flower10} with relatively high pre-shock densities . The \ho\, line ratios and upper limits for the other sources in the sample are overplotted (black points and arrows).}
  \label{fig:h2o_ratios2}
   \end{figure*}


Determining the origin of the observed molecular lines is more difficult because, as explained in Sect.~\ref{sect:molecular}, they are spectrally and spatially unresolved with PACS. The fact that the large CO and \ho\, line fluxes cannot be reproduced by disk models, with parameters typical of low mass YSO and T Tauri stars, favours either a shock origin or a PDR origin in UV-heated outflow cavities.
The molecular emission is more compact than the atomic emission but could, nonetheless, arise in a shock or an outflow. 

Evidence of molecular emission associated with a jet, but which is less extended than the atomic emission, has been found in evolved CTTSs, observed at near-infrared wavelengths. In particular, \citet{beck08} analysed high angular resolution observations ($\sim$0\farcs1) of the H$_2$ 2.12 \um\ line in CTTSs that are driving jets. They showed that the emission in this line is more compact than the emission in atomic optical forbidden lines (\oi, \sii, \azii) but still extends up to 1\arcsec-2\arcsec\, and is spatially associated with the jet direction.  \citet{beck08} investigated the origin of the H$_2$ emission in three of the sources analysed in this paper (T Tau, DG Tau A, and RW Aur). 
They detected molecular hydrogen at distances $\ge$50 AU from the star and derived excitation temperatures $>$1800 K. They also found that the emission lines were consistent with existing shock models. Based on these measurements and the kinematics of the features, they concluded that most of the H$_2$ toward these stars arose from shocks associated with the known HH objects rather than from quiescent disk gas illuminated by the central star.
Moreover, a few recent studies have shown that high excitation \ho\, lines and even higher-$J$ CO transitions (up to J = 46-45) can be produced in the outflows emanating from young Class 0 and I sources \citep{vankempen10,herczeg12}.

In view of the results above, we have attempted to simulate the observed molecular line emission by means of shock models.
As shown by \citet{hollenbach89}, fast J-type shocks give rise to a radiative precursor and are strongly dissociative; for a given \oi~63 \um\, line flux, the emission in the \ho\, lines is negligible.
Accordingly, we have compared the observed \ho\, and CO line ratios with the predictions of slow C-type (V$_{shock}$ = 10--40 \kms) and J-type (V$_{shock}$ = 10--30 \kms) shock models of \citet{flower10} (see Fig.~\ref{fig:h2o_ratios} and \ref{fig:h2o_ratios2}). In these shock waves, much of the mechanical energy is transformed into \ho\, and CO line radiation, as illustrated in Table 1 of \citet{flower10}. Figs.~\ref{fig:h2o_ratios} and \ref{fig:h2o_ratios2} suggest that relatively high pre-shock densities are required to reproduce the intensities of transitions from high excitation levels of \ho\,, such as the o-\ho\, 8$_{18}$ -- 7$_{07}$ line (E$_{\rm up}\sim$1070 K).

In the case of T Tau, all the \ho\, lines falling in the spectral range covered by our observations are detected with a good signal--to--noise ratio. A satisfactory fit to the observed line ratios is obtained for a C-type shock with a pre-shock density of 10$^6$ \cmc\, and a velocity in the range 20--30 \kms (see Fig.~\ref{fig:h2o_ratios2}). The diameter ($\sim$ 220--360~AU) of the emitting region that we estimate for T Tau is comparable with that deduced by \citet{spinoglio00}, using a large velocity gradient (LVG) model to fit the observed FIR lines.

For the other sources, we detected a maximum of four \ho\, lines. Under these circumstances, it makes little sense to look for a best--fit model;
but the few line ratios and the upper limits overplotted in Fig.~\ref{fig:h2o_ratios2} indicate that C- and J-type shocks with pre-shock densities of 10$^5$--10$^7$ \cmc\, and 10$^4$--10$^5$ \cmc, respectively, produce \ho\, line ratios which are consistent with those observed.
Moreover, the models can easily reproduce the large \ho\, 179.5 \um\, line fluxes for an emitting region with a diameter of a few tens to a few hundreds of AUs (see Tab.~\ref{tab:emitting_area}), which is consistent with the fact that the source of the molecular emission is compact and unresolved with PACS. 

The observed o-\ho\, 179.5 \um/CO 18-17 line ratios are lower, by up to a factor 4, than those predicted by the slow shock models: \ho/CO$_{\rm obs} \sim$ 0.2--2; \ho/CO$_{\rm shocks} \ge$ 0.8 (see right-hand panel of Fig.~\ref{fig:h2o_ratios}). Furthermore, the observed OH line fluxes are much larger (by up to a factor 10) than those predicted by these same models.
\ho\, abundances which are lower than predicted by slow shock models have been found by \citet{santangelo12} and \citet{vasta12}, from the analysis of a large number of \ho\, emission lines associated with outflows in Class 0 sources.
These authors propose that the water abundance may be reduced by UV dissociation \citep{bergin98} and/or depletion onto grains.
Photodissociation of \ho\, to OH, by the stellar FUV radiation field, is a possible explanation of these observations, as suggested also by \citet{spinoglio00}; but the further photodissociation of OH to O should also be taken into account. For further details, see Sect.~\ref{sect:iso_data}.

From the comparisons with the models, we conclude that the atomic and the molecular emission -- which have different spatial distributions (extended versus compact) -- arise in shock waves with different characteristics (C- or J-type, with or without a radiative precursor). This conclusion is consistent with the results of previous analyses of FIR ISO observations of Class 0 and I sources, which demonstrated that a single gas component cannot reproduce both the atomic and the molecular emission. It has been suggested that a dissociative J-type shock, occurring at the apex of the jet, and non-dissociative C-type shocks, occurring in the wings of the bow--like flow, may be  responsible for exciting the atomic and the molecular lines, respectively  \citep{nisini02}. Alternatively, the emission might arise from the UV-heated gas in the outflow cavity walls and small-scale C-type shocks occurring along the cavities \citep{vankempen10}.
We cannot exclude the possibility that the disk contributes to the molecular line emission, particularly to the higher excitation water lines, as suggested by a recent study \citep{riviere-marichalar12}.

Follow-up observations with HIFI and ALMA are planned, with the goal of resolving the molecular lines, spectrally (with HIFI) and spatially (with ALMA), and identifying the contributions to the emission from the disk, UV-heated outflow cavity walls, and shocks.



\begin{table}
\caption[]{\label{tab:emitting_area} Diameter of the \ho\, emitting area
 (in AU) estimated from C-type shock models.}
\vspace{0.2cm}
\centering
   \begin{tabular}[h]{c|c}
\hline\hline
source & R~(\ho) \\
          &  (AU)        \\
\hline
    T Tau     &   220 - 360 \\
 DG Tau A     &   35 - 270 \\
 DG Tau B     &   14 - 110 \\
FS Tau A+B     &   54 - 412 \\
   RW Aur     &   39 -298 \\
\hline

   \end{tabular}
\end{table}

\subsection{T Tau: comparison with previous ISO data}
\label{sect:iso_data}

  \begin{figure*}[!ht]
    \centering
    \includegraphics[width=8.8cm]{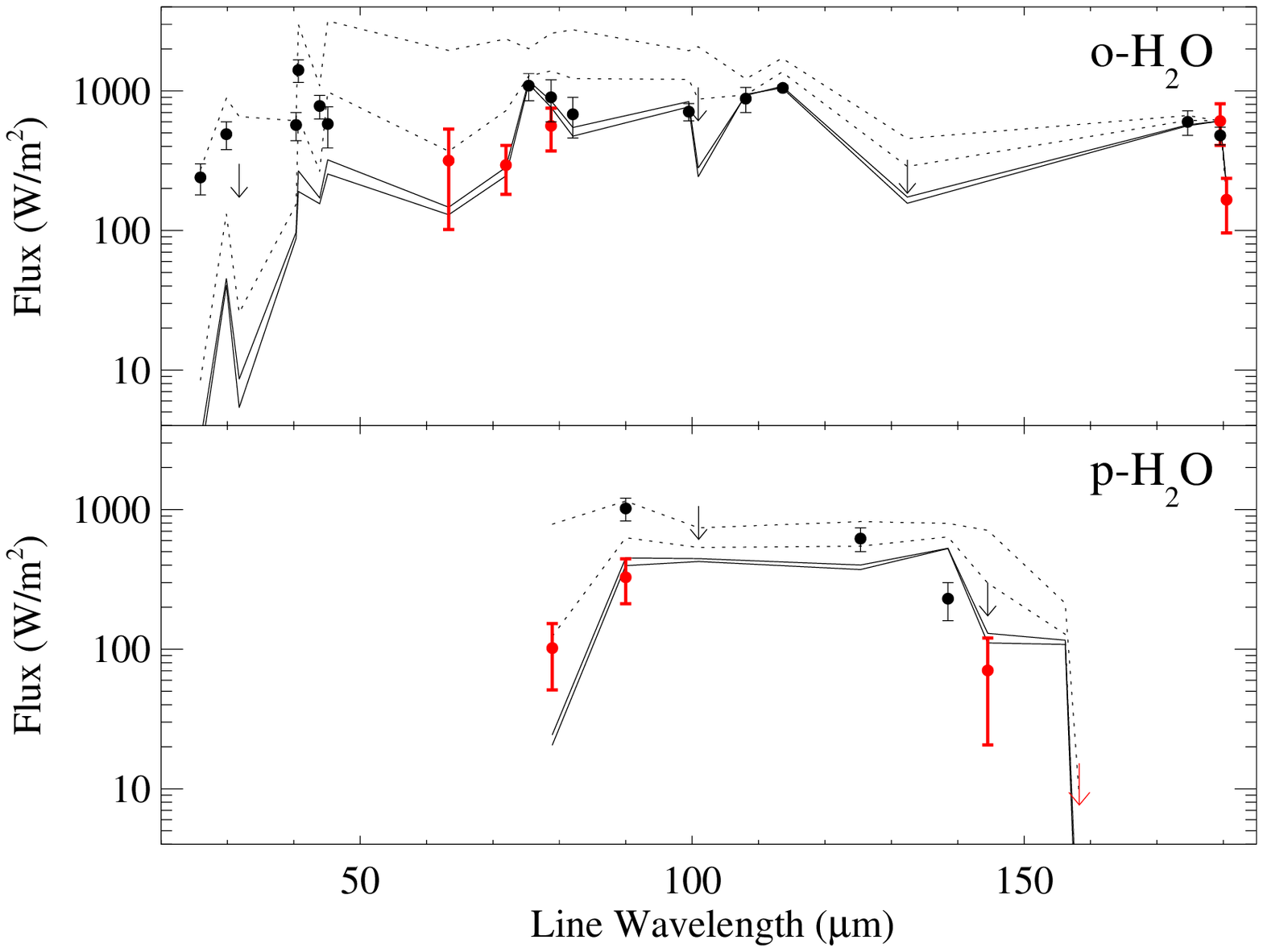}
    \includegraphics[width=8.8cm]{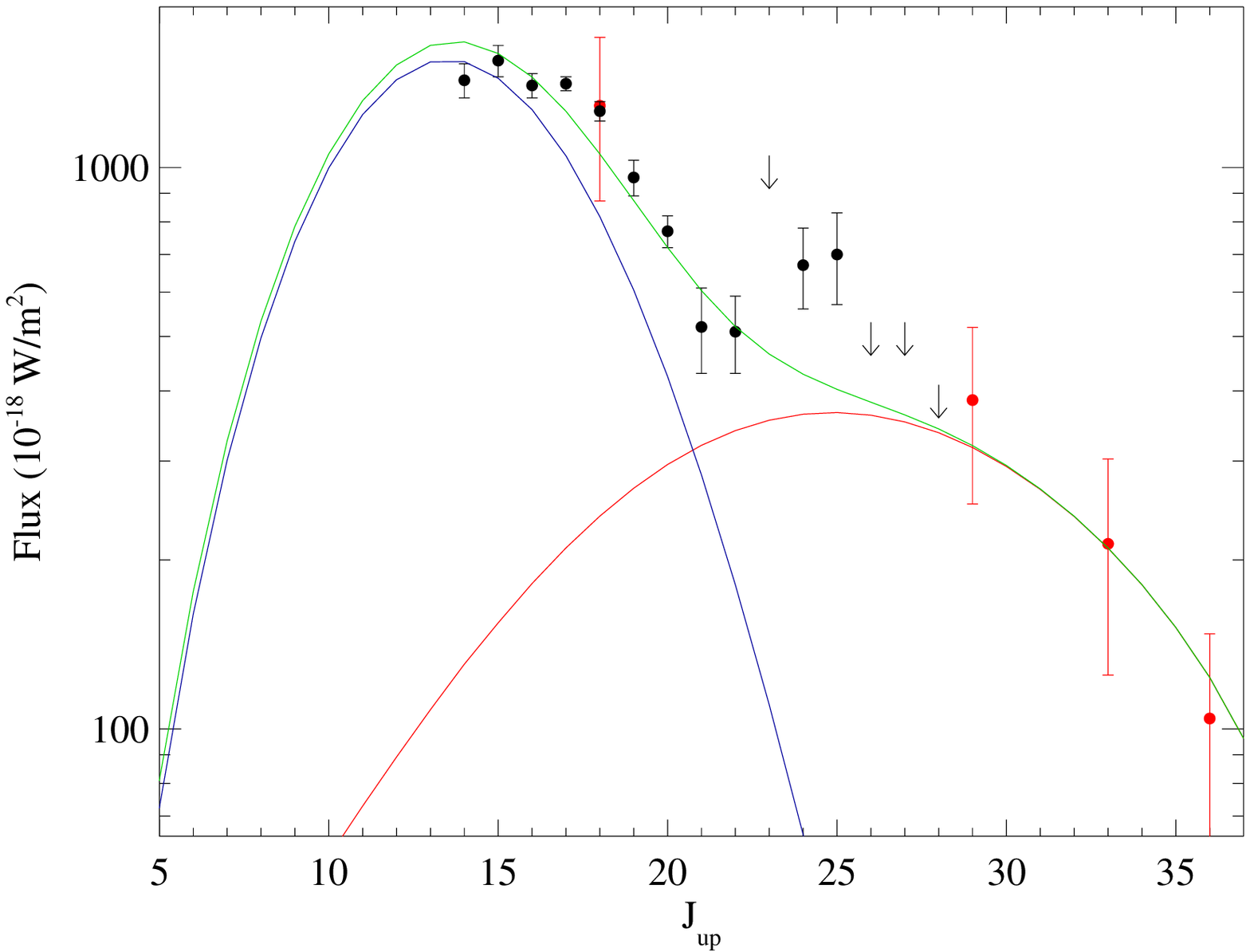}
 \caption{All the available ISO (black points) and {\it Herschel} (red points) observations of \ho\, ({\it left panel}) and CO ({\it right panel}) lines for T~Tau are shown. {\it Left panel:} Line fluxes predicted by a slow C-shock for an emitting area of $\sim$360 AU diameter and pre-shock density of n=10$^{6}$ \cmc\, (velocity V=20-30 \kms, dotted lines), or n=10$^{7}$ \cmc\, (V=10-20 \kms, dotted lines).
The low-excitation water lines are well reproduced by C-shock with pre-shock density of 10$^{6}$ \cmc\, while the higher excitation lines require higher pre-shock density. {\it Right panel:} CO line fluxes predicted by: a C-shock model (n=10$^{6}$ \cmc, V=20-30 \kms, emitting area diameter, D=360 AU, {\it red line}); an LVG warm gas component (n=10$^{6}$, T=300 K, D=360 AU, N$_{CO}$=1.5 10$^{18}$ cm$^{-2}$, {\it blue line}); the sum of the two models ({\it green line}). While the high-J CO lines are well reproduced by the same C-shock model used for water lines, a warm gas component is required to simultaneously fit the lower-J lines observed with ISO (down to J$_{up}$=14).}
  \label{fig:ttau_iso}
   \end{figure*}


T~Tau is the only source in our sample previously observed with ISO. The ISO 
observations were acquired with both the Short Wavelength Spectrometer (SWS) 
and the Long Wavelength Spectrometer (LWS) providing complete spectral 
coverage from 2~\um\, to 190~\um. These observations showed several \ho, CO 
(up to J$_{up}$=25), and OH emission lines \citep{vandenancker99,spinoglio00}. 
Our {\it Herschel}/PACS observations complement the previous ISO dataset by 
revealing  high-J CO lines (up to J=36-35) and \ho\, lines which were not 
detected with ISO because of its lower sensitivity.

The absolute line fluxes measured by ISO-LWS and {\it Herschel} are in very good 
agreement as shown in Fig.~\ref{fig:ttau_iso}. 
\citet{spinoglio00} modeled 
the observed lines by using an LVG code in a plane 
parallel geometry and found that most of the observed molecular emission 
may be explained by a dense and warm gas component (n=10$^5$-10$^6$ 
\cmc, T=300-900 K) with an emitting area of diameter of 300-400 AU.

Following these results we have modelled the full ISO+{\it Herschel} dataset with 
the shock models presented in the previous section.
The left panel of Fig.~\ref{fig:ttau_iso} shows that most of the water lines 
are well fit with a single C-shock model (n=10$^{6}$ \cmc, V$_{shock}$=20-30 \kms, 
diameter of the emitting area of $\sim$360 AU). The exceptions are for a few high 
excitation water lines between 25 and 50 \um\, observed with the ISO-SWS. 
To fit these lines, the pre-shock density must be about an order of magnitude larger (n=10$^{7}$ \cmc). 
However, this discrepancy between models and observations may be partially due to 
intercalibration problems between the SWS and LWS (the continuum 
flux measured by SWS is $\sim$10\% higher than in the LWS). 

The CO lines detected with ISO by \citet{spinoglio00} indicated that a warm component and a hot component are required
to reproduce all of the CO lines (Fig.~\ref{fig:ttau_iso}). This is further confirmed by our {\it Herschel}/PACS observations of CO lines up to J=36-35. While the very high-J CO lines detected by {\it Herschel} (CO J=36-35, 33-32, 29-28) are well fitted by the same C-shock model used to reproduce the \ho\, lines (we assumed that the \ho/CO abundance ratio is lowered by $\sim$0.55 due to FUV irradiation of the shocked region) a warm gas component is required to reproduce the lower-J CO lines (down to J$_{up}$=14). We tentatively fit all CO lines by adding a warm gas component using a RADEX LVG model in plane parallel geometry \citep{vandertak07} and choosing parameters which are similar to those used by \citet{spinoglio00} (n=10$^{6}$ \cmc, T=300 K, A=360 AU, N=1.5 10$^{18}$ cm$^{-2}$). This warm gas component also reproduces the OH line fluxes for an OH column density of $\sim$5 10$^{16}$ cm$^{-2}$.

More detailed modelling of all the detected lines could be made by considering 
UV-heated gas in the outflow cavity walls, and small scale C-shocks along them, 
as discussed by \citet{vankempen10}. However, such detailed modelling is beyond the scope of the present paper.

\subsection{FIR cooling: an evolutionary picture}
\label{sect:cooling}

\begin{table*}[!ht]
\begin{center}
\caption[]{\label{tab:FIR_cooling} Cooling in all the detected FIR lines.}
    \begin{tabular}[h]{cc|cccccc}
     \hline\hline
Source & Class & L~\oi & L~\cii & L~(OH) & L~(\ho) & L~(CO) & L~(FIR)$^{a}$ \\
            &           & (\lsol) & (\lsol) & (\lsol) & (\lsol) & (\lsol) & (\lsol) \\ 
\hline
     T Tau    & II + I &       1.2  10$^{-2}$ &        4.6  10$^{-4}$ &        1.3  10$^{-3}$ &        1.5  10$^{-3}$ -        1.6  10$^{-2}$ &        1.2  10$^{-3}$ -        9.7  10$^{-3}$ &        1.7  10$^{-2}$ -        4.0  10$^{-2}$  \\
  DG Tau A   & II &       1.1  10$^{-3}$ &        1.8  10$^{-4}$ &        4.9  10$^{-5}$ &        2.1  10$^{-5}$ -        4.2  10$^{-4}$ &        5.9  10$^{-5}$ -        2.5  10$^{-4}$ &        1.5  10$^{-3}$ -        2.0  10$^{-3}$  \\
  DG Tau B     &  I &      4.6  10$^{-4}$ &        1.5  10$^{-5}$ &        9.0  10$^{-6}$ &        2.7  10$^{-6}$ -        7.0  10$^{-5}$ &        8.0  10$^{-6}$ -        4.2  10$^{-5}$ &        4.9  10$^{-4}$ -        6.0  10$^{-4}$  \\
FS Tau A+B     & II + I &       3.3  10$^{-4}$ &        2.6  10$^{-5}$ &        2.2  10$^{-5}$ &        3.9  10$^{-5}$ -        9.7  10$^{-4}$ &        2.2  10$^{-5}$ -        5.8  10$^{-4}$ &        4.4  10$^{-4}$ -        1.9  10$^{-3}$ \\
    RW Aur     &  II &      1.3  10$^{-4}$ &        0.0  10$^{-5}$ &        1.5  10$^{-5}$ &        3.1  10$^{-5}$ -        5.1  10$^{-4}$ &        7.4  10$^{-6}$ -        3.0  10$^{-4}$ &        1.8  10$^{-4}$ -        9.5  10$^{-4}$  \\
\hline
\multicolumn{2}{c}{Class 0 $^{b}$} & 1 10$^{-3}$ - 4 10$^{-1}$ & 3 10$^{-4}$ - 2 10$^{-1}$ & 0                - 3 10$^{-1}$ & 0                 - 1.2              & 0 - 9 10$^{-1}$                 & 1 10$^{-3}$ - 2.8  \\
\multicolumn{2}{c}{Class I $^{c}$}  & 3 10$^{-4}$ - 1 10$^{-1}$ & 7 10$^{-4}$ - 1 10$^{-1}$ &  0                - 7 10$^{-3}$  & 0                 -  2 10$^{-2}$ &0 - 4 10$^{-2}$                 & 5 10$^{-4}$ - 1.4 10$^{-1}$ \\
\multicolumn{2}{c}{Class II $^{d}$} & 1 10$^{-4}$ - 1 10$^{-3}$ & 0                - 2 10$^{-4}$ &  1 10$^{-5}$ - 5 10$^{-5}$ & 4 10$^{-4}$ - 5 10$^{-4}$  & 2 10$^{-4}$ - 3 10$^{-4}$ & 9 10$^{-4}$ - 2 10$^{-3}$ \\    
\hline
        \end{tabular}
\end{center}
$^{a}$  L (FIR) =  L~\oi\, + L~(OH) + L~(\ho) + L~(CO)\\
$^{b}$ The luminosity values for Class 0 sources are from \citet{giannini01} and \citet{nisini10} (17 sources)\\
$^{c}$ The luminosity values for Class I sources are from this work (T Tau, DG Tau B, and FS Tau A+B), \citet{vankempen10} (HH 46), and \citet{nisini02} (14 sources) \\
$^{d}$ The luminosity values for Class II sources are from this work (DG Tau A, and RW Aur) (2 sources)\\
\end{table*}

\begin{table*}[!ht]
\begin{center}
\caption[]{\label{tab:mass_loss} Mass loss rates derived from the luminosity of the \oi~63~\um\, line are compared with \mjet\, values from optical forbidden lines and \macc\, estimates obtained from UV veiling and/or optical/NIR HI lines.}
    \begin{tabular}[h]{c|cccc}
     \hline\hline
Source  & \mjet~(\oi~63) & \mjet~(opt) $^{a}$ & \macc $^{b}$ & \mjet/\macc \\
            & (\msolyr) & (\msolyr) & (\msolyr) & \\ 
\hline
     T Tau     &       1.2  10$^{-6}$ &  1-7 10$^{-7}$ + ?                 & 0.3-1.5 10$^{-7}$ + ? & - \\
  DG Tau A     &       1.1  10$^{-7}$ &  3 10$^{-8}$ - 3 10$^{-7}$  & 0.5-2 10$^{-6}$ & 0.05 - 0.2 \\
  DG Tau B     &       4.4  10$^{-8}$ &  7 10$^{-9}$                         & 2.2 10$^{-7}$ & 0.2 \\
FS Tau A+B     &       3.1  10$^{-8}$ &  ? + $\le$2.5 10$^{-9}$       & 2-3 10$^{-7}$ & 0.17 \\
    RW Aur     &       1.2  10$^{-8}$ &  2 10$^{-9}$ - 2-3 10$^{-8}$ & 0.034-1.6 10$^{-6}$ & 0.008 - 0.35 \\
\hline
Class 0 $^{c}$ & 1~10$^{-7}$ - 4~10$^{-5}$  &  -                                       & - & - \\
Class I $^{d}$  & 3~10$^{-8}$ - 1~10$^{-5}$  &  10$^{-8}$ - 10$^{-6}$      & 10$^{-6}$ - 10$^{-5}$ & - \\
Class II $^{e}$ & 1~10$^{-8}$ - 1~10$^{-7}$   &  10$^{-11}$ - 3 10$^{-7}$ & 10$^{-10}$ - 4 10$^{-6}$ & - \\    
\hline
        \end{tabular}
\end{center}
$^{a}$ \mjet~(opt) estimated: from optical \oi, \sii\, line luminosity by \citet{white04} (T Tau N, FS Tau B, RW Aur), \citet{hartigan95} (DG Tau A, RW Aur), \citet{herbst97} (T Tau); and from jet density, velocity and radius estimates by \citet{coffey08} (DG Tau A), \citet{podio11} (DG Tau B), and \citet{melnikov09} (RW Aur). \\
$^{b}$ \macc\, estimates by \citet{hartigan95,hartigan03,gullbring00,white01,white04,calvet04,beck10}\\
$^{c}$ \mjet~(\oi~63) for Class 0 sources are from \citet{giannini01} (17 sources) \\
$^{d}$ \mjet~(\oi~63) for Class I sources are from this work (T Tau, DG Tau B, and FS Tau A+B), \citet{vankempen10} (HH 46), and \citet{nisini02} (14 sources),  \mjet~(opt) are from \citet{hartigan94,bacciotti99,podio06}, \macc\, from \citet{hartigan94}\\
$^{e}$ \mjet~(\oi~63) for Class II sources are from this work (DG Tau A, and RW Aur) (2 sources), \mjet~(opt) are from \citet{hartigan95,coffey08}, \macc\, from \citet{hartigan95,gullbring98}\\
\end{table*}



To evaluate the efficiency of the FIR outflow component associated with the jet sources in our sample we estimate the total luminosity radiated away in the far-infrared lines, L~(FIR).

The \oi, \cii, and OH luminosities were estimated from the observed line fluxes (L~\oi = L (\oi\,63 + 145 \um),  L~\cii\, = L~\cii\, 157 \um, L~(OH) = L (OH 79.11 + 79.18 \um)) while for the CO and \ho\, luminosity we give a lower limit, inferred by summing the fluxes of the observed lines, and an upper limit inferred by adding up the predicted line intensities for all the transitions considered in the shock models (i.e. 45 levels of o-\ho,  45 levels of p-\ho,  and 41 levels of CO, \citealt{flower10}).
In particular, we consider the highest luminosity value obtained from the different shock models which reproduce the observed line ratios (i.e., J-shocks with pre-shock densities of n$_0$=10$^{4}$-10$^{5}$ \cmc\, and C-shocks with   n$_0$=10$^{5}$-10$^{7}$ \cmc).
Then the cooling in the molecular lines is computed as L$_{mol}$ = L~(OH) + L~(\ho) + L~(CO), while the total cooling in FIR lines is estimated by summing up the cooling in all the detected species: L (FIR) =  L~\oi\, + L~(OH) + L~(\ho) + L~(CO). Note that following previous ISO studies \citep{giannini01, nisini02} we neglect the luminosity of the \cii\, line when computing L~(FIR) because the emission in this line may be contaminated by cloud emission. 
%
The estimated values of cooling in all the observed species (\oi, \cii, OH, \ho, CO) and the total FIR line cooling are summarised in Tab.~\ref{tab:FIR_cooling}. 

The line cooling at different evolutionary stages is shown in Fig.~\ref{fig:line_cooling} by means of histograms of L~\oi, L$_{mol}$, and L~(FIR)/L$_{bol}$ for Class 0, I, and II sources.
The luminosity values reported in the figure are for: the Class 0 sources observed with ISO by \citet{giannini01} and with {\it Herschel} by \citet{nisini10} (17 Class 0 sources); the Class I (or unresolved Class II + I) sources analysed in this paper (T Tau, DG Tau B,  and FS Tau A+B), complemented by ISO observations of Class I sources by \citet{nisini02}, and PACS observations of HH 46 by \citet{vankempen10} (14 Class I sources); the Class II sources analysed in this work (DG Tau A and RW Aur) (2 Class II sources).  
Note that  only 16 out of 17 Class 0 sources and 11 out of 14 Class I sources are reported in the histogram of the molecular cooling (middle panel of Fig. ~\ref{fig:line_cooling}). The other sources in the sample by \citet{giannini01} and \citet{nisini02} do not show molecular line emission, probably due to the limited sensitivity of ISO.

Despite the small statistical samples, in particular for Class II sources, the histograms in Fig. ~\ref{fig:line_cooling} and the range of values reported in Tab.~\ref{tab:FIR_cooling} indicate that the total FIR cooling decreases with the source evolutionary stage going from values of  $\sim$10$^{-3}$ - 3 \lsol\, for Class 0 sources to values of  $\sim 5~10^{-4}$ - $10^{-1}$  \lsol\, and $\sim 9~10^{-4}$ - $2~10^{-3}$ \lsol\, for Class I and Class II sources.
In particular,  L~\oi\, is of $\sim 10^{-3} - 4~10^{-1}$ \lsol\, in Class 0 sources, of $\sim 3~10^{-4} - 10^{-1}$ \lsol\, in Class I, and  $\sim 10^{-4} - 10^{-3}$ \lsol\, in Class II sources. The cooling in the molecular lines, L$_{mol}$, shows a stronger decrease going from Class 0 (L$_{mol}$ up to 2.4 \lsol) to Class I (L$_{mol}$ up to 0.04 \lsol), and Class II sources (L$_{mol} \sim 7-8~10^{-4}$ \lsol). This is due to progressive clearing of the circumstellar material which is accreted or transported away by the observed jets.
Finally, the bottom panel of Fig.~\ref{fig:line_cooling} shows that also L~(FIR)/L$_{bol}$ is decreasing going from Class 0, to Class I, and II sources indicating that the outflow efficiency in radiating away the source bolometric luminosity is decreasing with its evolutionary state. 
Similar results are obtained by Karska et al. ({\it in preparation}) for a sample of Class 0 and I sources observed with {\it Herschel}/PACS.

  \begin{figure}[!ht]
    \centering
    \includegraphics[width=7.cm]{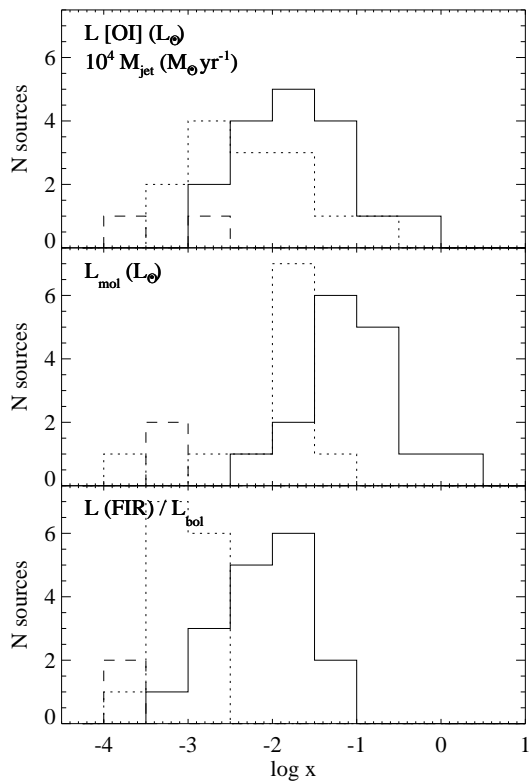}
 \caption{Histograms of L~\oi\, ({\it upper panel}), L$_{mol}$ ({\it middle panel}), and L~(FIR)/L$_{bol}$ ({\it bottom panel}) for Class 0, I, and II sources (solid, dotted, and dashed lines).}
  \label{fig:line_cooling}
   \end{figure}

\subsection{Mass loss rates}
\label{sect:mloss}

The mass loss rate, \mjet, is estimated from the \oi\, 63 \um\, luminosity by using the relationship by \citet{hollenbach85}, from which it is shown that if the ejected material is moving fast enough to produce a dissociative J-shock, then \oi\, emission will be the dominant coolant in the postshock gas for temperatures of 100-5000 K. Thus, the \oi\, luminosity is a direct tracer of the mass flow into the shock, and hence of the mass loss rate, \mjet:    

\be
$\mjet$ ($\msolyr$) = 10^{-4} L~$\oi$ 63 $\um$ ($\lsol$) 
\ee

This is a simpler method to estimate \mjet\, than the use of optical lines \citep[e.g., ][]{hartigan95,bacciotti99,podio06} because it does not depend on estimates of the visual extinction, the inclination of the system, or the geometry of the outflow. 
On the other hand, the derived estimates are based on the assumption that all the ejected material is J-shocked and that all the observed \oi\, emission is produced by shocks. Thus, if part of the observed \oi~63~\um\, emission arises from a photodissociation region due to, e.g., the UV-illuminated outflow cavities, and/or from the disk we may overestimate the mass loss rate.

The mass loss rates derived from the \oi\, 63 \um\, line are summarised in Table \ref{tab:mass_loss} and compared with mass loss and mass accretion rates derived from optical lines and UV veiling \citep[e.g., ][]{hartigan95,gullbring98,gullbring00}.  The mass loss rates derived from the  \oi\, 63 \um\, line are also compared with the values estimated previously for Class 0 and I sources \citep{giannini01, nisini02} (see top panel of Fig.~\ref{fig:line_cooling}).
In agreement with the estimated FIR cooling, the mass outflow rates derived from the \oi\, 63 \um\, line decrease as the driving source evolves from values of $\sim 1~10^{-7} - 4~10^{-5}$ \msolyr\, for Class 0, to $\sim 3~10^{-8} - 1~10^{-5}$ \msolyr\, for Class I, and down to $\sim 10^{-8} - 10^{-7}$ \msolyr\, for Class II sources.

The comparison of \mjet(\oi~63~\um) with \mjet\, and \macc\, values derived from observations at optical and UV wavelengths is difficult because the latter show discrepancies up to one order of magnitude depending on the adopted method. This is mainly due to the fact that \mjet\, and \macc\, values derived from optical lines and UV veiling are highly dependent on the estimates of the jet radius and visual extinction (see, e.g., \citealt{hartigan95,bacciotti99,podio06} for a discussion of different methods to derive \mjet\, from optical forbidden lines and \citealt{gullbring98} for a discussion of the uncertainties affecting \macc\, estimates).  Thus, for \mjet~(opt) and \macc\, estimates we report in Table~\ref{tab:mass_loss} a range covering all the different values found in the literature. However, both for Class II and Class I sources the mass loss rate derived from the \oi~63~\um\, line luminosity is larger than or comparable to the maximum value obtained from optical and NIR forbidden lines. This suggests that the mass loss rate can actually be larger than previously thought and the ejection rate can be up to a few percent of the accretion rate (\mjet/\macc\, up to 0.35). This is even more evident if we consider that the lower \mjet/\macc\, ratios are derived by using the high \macc\, values estimated by \citet{hartigan95}. More recent work by \citet{gullbring98} showed that these values can be overestimated by one to two orders of magnitude. \citet{cabrit07} found similar high \mjet/\macc\,  when considering the most recent and accurate \macc\, and \mjet\, estimates and showed that high \mjet/\macc\, values may have important implications for jet launching models. For example, they show that stellar winds cannot produce such high mass loss rates while X- \citep{shu94} and Disk- \citep{ferreira06} wind models may provide mass ejection to mass accretion ratios up to 0.1-0.25.  

Another fundamental issue is to understand how the mass ejection to mass accretion ratio evolves with the source evolutionary state. The values derived for the Class I and II sources in our sample seems to suggest that the \mjet/\macc\, ratio remains constant. However, it is not possible to draw a firm conclusion given the large uncertainties affecting the mass accretion rate estimates and the small size of the considered sample.


\section{Conclusions}
\label{sect:conclusions}

In this paper we have analysed {\it Herschel}/PACS integral-field spectroscopic observations of Class I and II sources in Taurus which are known to drive bright optical jets. 
Thanks to the {\it Herschel} sensitivity (100-1000 times larger than ISO) we are able to detect the FIR counterpart of optical jets from the selected Class I and II sources  for the first time.
An exception is T Tau which is a bright multiple system unresolved with PACS, consisting of a Class II source and a Class I binary system, and associated with at least two jets, which has been observed with ISO \citep{spinoglio00}.
We investigate the origin of the detected atomic and molecular lines by carefully evaluating the spatial distribution of the emission on the PACS detector and by comparing line fluxes and ratios with predictions from disk and shock models.
The results of our analysis are summarised below:

\begin{itemize}

\item[-] {\it the emission in the atomic \oi\, and \cii\, lines is extended and spatially correlated with the optical jet emission}. 
In two cases (DG Tau B and RW Aur) we also detect a consistent offset in velocity in all the spaxels where the \oi~63~\um\, line is detected which indicates a gas velocity in agreement with the values measured for the associated optical jets.   \\ 

\item[-] {\it the emission in the molecular \ho, CO, and OH lines is spatially and spectrally unresolved}. 
However, by using the DENT grid of models we show that for typical low mass YSO and T Tauri star parameters the irradiated disk surface is unlikely to produce the observed large \ho, CO fluxes (up to 10$^{-16}$ and 10$^{-15}$ \wm, respectively) even when the source is associated with a strong X-ray field. Slow C- and J- shocks (V$_{shock} \le$40 \kms\, and V$_{shock} \le$30 \kms, respectively), on the other hand, can reproduce the observed line fluxes for an emitting area of diameter of a few tens to a few hundreds of AU. Thus, a shock origin is favoured.\\

\item[-] {\it high-J CO lines (up to CO J=36-35) and \ho\, lines from high excitation levels (up to E$_{up}\sim$1070 K) are detected} similarly to what was observed by \citet{vankempen10} and \citet{herczeg12} for Class 0 and Class I outflow sources (NGC 1333 IRAS 4B and HH 46/47, respectively). This suggests that lines from high excitation levels can be shock excited if the density is high enough.\\

\item[-] {\it the extended atomic emission may be produced by fast J-shocks}. Shocks with velocities higher than 30 \kms\, with a radiative precursor \citep{hollenbach89} strongly dissociate and ionize the gas giving rise to high \oi\, and \cii\, line fluxes, in agreement with the observed line ratios (\oi\, 63/145$\sim$15-30, \cii / \oi$\le$0.17). Excess \cii\, emission may be due to UV-heated gas in the outflow cavity walls.\\

\item[-] {\it molecular emission may originate instead in slow C- or J- shocks}, which preserve molecules \citep{flower10}. High pre-shock densities are required to populate the high excitation \ho\, levels and reproduce the observed line ratios (i.e. J-shocks with n$ \sim$10$^4$-10$^5$ \cmc\, or C-shocks with n$ \sim$10$^6$-10$^7$ \cmc). We cannot exclude, however, that the disk and the warm gas in the outflow cavity walls are contributing to the observed emission. \\

\item[-] {\it the cooling in the FIR lines is decreasing as the source evolves:} for the Class II sources in our sample the cooling is from one to four orders of magnitude lower than for Class I and 0 sources (L~\oi\,$\sim$10$^{-4}$--10$^{-3}$ \lsol, L~\ho\, and L~CO $\sim$10$^{-4}$ \lsol). The molecular cooling is decreasing more abruptly as the source evolves indicating that for Class 0 sources the main coolants are water and CO, while in Class I and II \oi\, becomes an important coolant. \\

\item[-] {\it the mass loss rate for the Class II sources in our sample is  up to three orders of magnitude lower than for Class 0 and I sources}, i.e. \mjet (\oi~63~\um) $\sim 10^{-8} - 10^{-7}$ \msolyr. \\

\item[-] {\it the mass loss rates inferred from the \oi~63~um\, line are larger than or comparable to values obtained from optical and NIR forbidden lines,} implying higher mass ejection to mass accretion ratios, up to 0.35. This may have important implication for jet launching models. \\

\end{itemize}

The above summary places the FIR emission from Class II and I jet sources within an evolutionary picture.
The Taurus optical-jet-sources studied in this work show FIR atomic and molecular emission similar to that previously observed with ISO for Class 0 and Class I sources, including a highly excited molecular component.
However, the emission associated with Class II sources is fainter and more compact (in particular the molecular component), and the FIR line cooling and mass loss rates are one to three orders of magnitude lower than those estimated for Class 0 and I sources.

\begin{acknowledgements}
L. Podio acknowledges the funding from the FP7 Intra-European Marie Curie Fellowship (PIEF-GA-2009-253896). C. Howard, G. Sandell, and S. Brittain acknowledge NASA/JPL. I. Kamp acknowledges funding from an NWO MEERVOUD grant. We thank the PACS instrument team for their dedicated support. We also thank J. Eisl\"offel for letting us use the images published in \citet{eisloffel98} for the plots in Fig.~\ref{fig:OI_maps}. Finally we are grateful to B. Nisini, C. Codella, A. Karska and the referee for very useful discussion and comments that helped improving the paper.          
\end{acknowledgements}

\bibliographystyle{aa} 
\bibliography{mybibtex} 

\appendix

\section{{\it Herschel}/PACS observation identifiers}
\label{app:obs}

Table \ref{tab:obsid} list the targets, observational modes, operational days (ODs), and identifiers (OBSIDs) of the {\it Herschel}/PACS observations analysed in the paper.

\begin{table}
  \caption[]{\label{tab:obsid} Target, observational mode (LineSpec or RangeSpec), operational day (OD), and identifier (OBSID) of the analysed {\it Herschel}/PACS observations.}
\centering
\begin{tabular}[h]{cccc}
\hline\hline
  Target     & Obs. Mode  & OD & OBSID \\
\hline
T Tau         & LineSpec     & 272 & 1342190353 \\
                  & RangeSpec  & 272 & 1342190352 \\
DG Tau A   & LineSpec     & 273 & 1342190382 \\
                  & RangeSpec  & 273 & 1342190383 \\
DG Tau B    & LineSpec     & 316 & 1342192798 \\
                  & RangeSpec  & 678 & 1342216652 \\
FS Tau A+B & LineSpec     & 316 & 1342192791 \\
                  & RangeSpec   & 641 & 1342214358 \\
RW Aur       & LineSpec      & 290 & 1342191359 \\
                  & RangeSpec   & 290 & 1342191358 \\
 \hline 
 \end{tabular}
\end{table}

\section{Line extended emission characterisation}
\label{app:extended_emission}

In this Appendix we describe the procedure which is applied to ascertain whether the line emission is extended and/or offset with respect to the continuum emission. 
If both line and continuum emission originate from the same region, supposedly the star-disk system which is unresolved with PACS, any emission detected out of the central spaxel is due to the spectroscopic PSF and/or to the fact that the source is not centred on the central spaxel.
The line and the continuum PSF have the same shape and are centred at the same position, hence the line-to-continuum ratio is constant across the PACS field  of view.
If, on the contrary, the line emitting region is more extended and/or offset with respect to the continuum emitting region the line and continuum distribution across the PACS field of view are different and the line-to-continuum ratio should vary in the different spaxels.
However, when most of the line emission is emitted close to the source it can be difficult to detect extended and/or offset emission given the low spatial resolution and sampling offered by PACS.

To check for the presence of extended line emission we subtract from the line + continuum image the ``on-source'' emission and search for ``residual'' local emission above the confidence level.

An image of the line and continuum emission is constructed integrating the PACS data cube over the wavelength range covered by the considered line. 
For example to obtain line+continuum image for the \oi~63~\um\, line we integrated on the spectral range from 63.141~\um\, to 63.242~\um, for all the sources in our sample. 
Integrated fluxes, F$_{line+cont}$, and errors, $\Delta$F$_{line+cont}$, in units of \wm, have been computed for each spaxel summing the flux density on each spectral element, $f_i$, in Jy, over the $n$ spectral elements in the defined wavelength range as follows:

\be
 F_{line+cont}  = d\nu \sum\limits_{i=1}^n f_i
\ee
\be
 \Delta F_{line+cont}  = d\nu \sqrt{ \sum\limits_{i=1}^n \Delta f_i^{2} }	
\ee

where d$\nu$ is the average spectral element size, in Hz. 
The continuum image, F$_{cont}$, with the associated error, $\Delta$F$_{cont}$,
in Jy, are estimated in each spaxel by computing a weighted average of
the flux density over a region of $n1$ spectral elements adjacent to the detected line:

\be
w_i = \frac{1}{\Delta f_i^2}
\ee

\be
F_{cont}  = \sum\limits_{i=1}^{n1}  \frac{f_i w_i}{\sum\limits_{j=1}^{n1} w_j}	
\ee

\be
\Delta F_{cont}  = \frac{1}{\sqrt{\sum\limits_{i=1}^{n1} w_i}} = \frac{1}{\sqrt{\sum\limits_{i=1}^{n1} \frac{1}{\Delta f_i^2}}}	
\ee

Then we subtract from the line + continuum image, F$_{line+cont}$, the ``on-source'' emission, i.e. the line + continuum flux in the spaxel showing the brightest continuum, which is scaled according to the continuum level in each spaxel.
If $j$ is the spaxel where the continuum emission is maximum, we obtain the image of the residual flux, F$_{residual}$, and the associated error, $\Delta$F$_{residual}$, as:

\be
 F_{residual}  = F_{line+cont} - \frac{F_{cont}}{R_j}
\ee
\be
R_j = \left( \frac{F_{cont, j}}{F_{line+cont, j}} \right) 
\ee
\be
\Delta~R_j = R_j \sqrt{ \left( \frac{\Delta F_{line+cont, j}}{F_{line+cont, j}} \right)^{2} 
+ \left( \frac{\Delta F_{cont, j}}{F_{cont, j}} \right)^{2}  } 
\ee
\be
\Delta F_{residual}  = \sqrt{  \Delta F_{line+cont}^2 + \frac{F_{cont}^2}{R_j^2}
\left( \frac{\Delta R_j^2}{R_j^2} + \frac{\Delta F_{cont}^2}{F_{cont}^2} \right)  }
\ee

The confidence level at which residual emission is detected in each spaxels is:

\be
\sigma = \frac{F_{residual}}{\Delta F_{residual}} 
\ee

Fig.~\ref{fig:OI_maps} (right panels) shows the displacement of the \oi\, 63 \um\, residual emission,  F$_{residual}$, with respect to the continuum emission, F$_{cont}$, and the optical jet direction.
We clearly see an offset between the continuum emission (dotted contours) and the \oi~63~\um\, line residual emission (solid contours) which is displaced along the optical jet PA (blue/red dashed lines). The spatial correlation between the \oi~63~\um\, line residual emission and the optical jet is evident also for the sources for which the line and continuum emission peak on the same spaxel before the subtraction of ``on-source'' emission (e.g., T Tau and RW Aur; see, for comparison the maps in the left panels of Fig.~\ref{fig:OI_maps}).


\end{document}